\documentclass[11pt]{article}
\usepackage{amssymb}
\title{On integrable Hamiltonians for higher spin $XXZ$ chain}
\author{\sc Andrei G. Bytsko
 \thanks{Alexander von Humboldt Fellow,
 on leave from Steklov Mathematics Institute, St.Petersburg, Russia
 \newline \indent\  bytsko@physik.fu-berlin.de} \\ [2mm]
 Institut f\"ur Theoretische Physik \\
 Freie Universit\"at Berlin\\
 Arnimallee 14, 14195 Berlin, Germany}
\date{ \small December, 2001 \\  \small hep-th/0112163}
\begin{document}
\maketitle
\begin{abstract}
\noindent
 Integrable Hamiltonians for higher spin periodic 
 $XXZ$ chains are constructed in terms of the spin 
 generators; explicit examples for spins up to 
 $\frac 32$ are given. Relations between Hamiltonians 
 for some $U_q(sl_2)$-symmetric and $U(1)$-symmetric 
 universal $r$-matrices are studied; their properties
 are investigated. A certain modification of the
 higher spin periodic chain Hamiltonian is shown
 to be an integrable $U_q(sl_2)$-symmetric
 Hamiltonian for an open chain. 
 
\end{abstract}
\section{Introduction}
$XXZ$ spin chains have numerous connections with 
two-dimensional statistical physics and  
(1+1)-dimensional quantum field theory.
They describe interaction of $q$-deformed spins sitting 
at the nodes of a one-dimensional lattice. 
The spin generators $S^+$, $S^-$ and $S^3$ obey 
the commutation relations of the quantum Lie algebra 
$U_q(sl_2)$ \cite{KR}
\begin{equation}\label{cS}
  [S^+ , S^- ] = \frac{\sin(2\gamma S^3)}{\sin\gamma} \,,\qquad
  [S^3 , S^\pm  ] = \pm S^\pm \,.
\end{equation}

We will consider only integrable $XXZ$ spin models. 
The simplest example in this class is a spin-$\frac 12$
chain with the Hamiltonian given by
\[  
  H = \sum_{n} \Bigl( \frac 12 (S^+_n S^-_{n+1} + 
 S^-_n S^+_{n+1}) + (\cos\gamma)\, S^3_n S^3_{n+1} \Bigr) \,.
\]
Higher spin $XXZ$ chains have also been studied 
\cite{FZ,XXZ,TTF} but only in the spin-1 case
an explicit expression for the corresponding Hamiltonian 
has been found \cite{FZ}. 

In the present paper we will construct local integrable 
Hamiltonians (in terms of the spin generators) for higher 
spin $XXZ$ chains. In Section~\ref{uR} we recall some 
facts about $U_q(sl_2)$-symmetric universal $r$-matrix 
$r(\lambda)$. In Section~\ref{sH} we construct 
a~$U_q(sl_2)$-symmetric local Hamiltonian~$H_{n,n+1}$. 
Its properties and properties of the corresponding closed 
chain Hamiltonian~$\cal H$ are discussed in Section~\ref{pR}.
In particular, we observe that $H_{n,n+1}$ decomposes
into a {$U(1)$-symmetric} bulk Hamiltonian 
$\widehat{H}_{n,n+1}$ plus a universal local boundary term.
Section~\ref{eX} contains explicit expressions for $H_{n,n+1}$
for spins $\frac 12$, 1 and $\frac 32$.
In Section~\ref{tC} we find a family of universal 
$r$-matrices and Hamiltonians corresponding to an alternative
choice of the co-multiplication. In Section~\ref{tL} we 
construct another family of ($U(1)$-symmetric) universal 
$r$-matrices and local Hamiltonians which contains
the Hamiltonian $\widehat{H}_{n,n+1}$ and the corresponding 
reflection-symmetric universal $r$-matrix $r_0(\lambda)$. 
In Section~\ref{oC} we employ our construction to
establish the integrability of certain 
($U_q(sl_2)$-symmetric) Hamiltonians for an open chain.
Appendix~A contains technical details related to $r(\lambda)$
and $r_0(\lambda)$. Appendix~B provides some details
on computation of the spin-1 and spin-$\frac 32$
Hamiltonians presented in Section~\ref{eX}. Appendix~C
explains a $q$-trace formula for $H_{n,n+1}$ used
in Section~\ref{oC}.

For compactness of notations, we will use in the text
both the deformation parameter~$\gamma$ introduced in 
(\ref{cS}) and~$q \equiv e^{{\rm i} \gamma}$.
We assume that $q$ is either real or takes values on the
unit circle. In the latter case $q$ is assumed to be
generic, i.e., it is not a root of unity.

\section{Universal $r$-matrix} \label{uR}
The starting point of the quantum inverse scattering 
method is the exchange relation
\begin{equation}\label{RLL}
 R(\lambda) \,  L(\lambda+\mu) \otimes L(\mu) =
 L(\mu) \otimes L(\lambda+\mu) \, R(\lambda) \,,
\end{equation}
with $\otimes$ understood as the tensor product with respect 
to an auxiliary space $V$ (below it is the space of
2$\times$2 matrices) and the usual product of operators 
in the quantum space $\mathfrak{H}$. The $R$-matrix 
belongs to $V\otimes V$. Thus (\ref{RLL}) is an equation
in $V\otimes V \otimes \mathfrak{H}$.

The following $L$-operator, i.e., an element of 
$V \otimes \mathfrak{H}$, is consistent with the 
algebra (\ref{cS})
\begin{equation}\label{LS}
  L(\lambda) = \frac{1}{\sin\gamma} \left( 
 \begin{array}{cc} \sinh [\gamma(\lambda + {\rm i} S^3)] &
  {\rm i}\, \sin\gamma \, e^{\gamma\lambda} \, S^-  \\ 
 {\rm i}\, \sin\gamma \, e^{-\gamma\lambda} \, S^+ &
 \sinh [\gamma(\lambda - {\rm i} S^3)] 
 \end{array} \right) 
\end{equation}
in the sense that it satisfies eq.~(\ref{RLL}) provided
that $R(\lambda)= P_V \check{R}(\lambda)$, where 
\begin{equation}\label{R}
 \check{R}(\lambda) =  {\rm i}\, e^{\gamma\lambda} \,
 \sigma^+ \otimes \sigma^- + {\rm i}\, e^{-\gamma\lambda} \,
 \sigma^- \otimes \sigma^+  
   + \frac{1}{\sin\gamma}  \sinh \Bigl( \gamma\lambda +
 \frac{{\rm i}\,\gamma}{2} (1 \otimes 1 + 
   \sigma^3 \otimes \sigma^3) 
 \Bigr) \,.
\end{equation}
Here $\sigma^a$ denote the Pauli matrices,
$\sigma^\pm = \sigma^1 \pm {\rm i} \sigma^2$, and 
$P_V=\frac 12 (1 \otimes 1 + \sum_a \sigma^a \otimes \sigma^a)$
is the permutation matrix in $V \otimes V$.

The $L$-operator (\ref{LS}) decomposes into 
two $\lambda$-independent Borel components,
\begin{equation}\label{Bor}
  (2 \sin\gamma) \, L(\lambda) = 
 e^{\gamma\lambda} L_+ - e^{-\gamma\lambda} L_- \,,
\end{equation}
which can be utilized to define \cite{FRT} the 
co-multiplication (a linear homomorphism 
$\Delta: U_q(sl_2) \rightarrow [U_q(sl_2)]^{\otimes 2}$)
in the matrix form,
\begin{equation}\label{Ldel}
  \Delta (L_\pm) =  L_\pm {\dot\otimes} L_\pm \,,
\end{equation}
where $\Delta$ acts on the quantum space and $\dot\otimes$ 
denotes the tensor product with respect to $\mathfrak{H}$ and the 
usual matrix product in $V$. In explicit form (\ref{Ldel}) reads
\begin{equation}\label{del}
 \Delta(S^\pm) = S^\pm \otimes q^{-S^3} + q^{S^3} \otimes S^\pm \,,
 \qquad \Delta(S^3) = S^3 \otimes 1 + 1 \otimes S^3 \,.
\end{equation}

Observe that (\ref{R}) can be obtained by evaluating
the quantum space of $L(\lambda + \frac{\rm i}{2})$ 
in the fundamental representation, where 
$S^\pm = q^{\pm \frac 12} \sigma^\pm$ and $S^3= \frac 12 \sigma^3$.
In fact, $\check{R}(\lambda)$ and $L(\lambda)$
are, respectively, $V \otimes V$ and $V \otimes \mathfrak{H}$
representations of a more general object, 
\hbox{${\bf \check{R}}(\lambda) \in [U_q(sl_2)]^{\otimes 2}$},
that satisfies the Yang-Baxter equation in
$\mathfrak{H} \otimes \mathfrak{H} \otimes \mathfrak{H}$,
\begin{equation}\label{RRR}
 {\bf \check{R}}_{12}(\lambda) \, 
 {\bf \check{R}}_{13}(\lambda+\mu) \,
 {\bf \check{R}}_{23}(\mu) =
 {\bf \check{R}}_{23}(\mu) \, 
 {\bf \check{R}}_{13}(\lambda+\mu) \, 
 {\bf \check{R}}_{12}(\lambda) \,.
\end{equation}
Relation (\ref{RLL}) is then a particular case of (\ref{RRR}) 
with the first and second spaces evaluated in the fundamental
representation. Evaluating (\ref{RRR}) in the 
$V \otimes \mathfrak{H} \otimes \mathfrak{H}$ representation
(and interchanging $\lambda$ with $\mu$), we obtain
\begin{equation}\label{rLL}
 r(\lambda) \,  L(\lambda+\mu) \dot\otimes L(\mu) =
 L(\mu) \dot\otimes L(\lambda+\mu) \, r(\lambda) \,,
\end{equation}
with $r(\lambda) \equiv {\bf \check{R}} (\lambda) {\bf P}$,
where ${\bf P}$ is the permutation in 
$\mathfrak{H} \otimes \mathfrak{H}$.
For a given \hbox{$L$-operator}, (\ref{rLL}) can be regarded 
as a defining equation on the element 
$r(\lambda) \in [U_q(sl_2)]^{\otimes 2}$ which we will call 
the universal $r$-matrix. Using (\ref{Bor}), equation 
(\ref{rLL}) can be equivalently rewritten as follows
\begin{eqnarray}
 \label{rllpm}
 & r(\lambda) \  L_\pm \dot\otimes L_\pm =
 L_\pm \dot\otimes L_\pm \ r(\lambda) \,, & \\
 \label{rll0}
 & r(\lambda) \, \bigl(e^{\gamma\lambda} L_+ \dot\otimes L_- +
 e^{-\gamma\lambda} L_- \dot\otimes L_+ \bigr) = 
 \bigl(e^{-\gamma\lambda} L_+ \dot\otimes L_- +
 e^{\gamma\lambda} L_- \dot\otimes L_+\bigr) \, r(\lambda) \,. &
\end{eqnarray}
In view of (\ref{Ldel}), the first line implies that 
for any element $\xi\in U_q(sl_2)$ we have
\begin{equation}\label{rd}
 r(\lambda) \, \Delta(\xi) = \Delta(\xi) \, r(\lambda) \,.
\end{equation}

Recall that, for a generic $q$, the tensor product of two 
irreducible highest weight $U_q(sl_2)$-representations of spin 
$S$ is completely reducible and decomposes into the sum 
\begin{equation}\label{tpr}
D_S \otimes D_S = \sum_{j=0}^{2S} D_j \,,
\end{equation}
where each subspace $D_j$ is a highest weight $U_q(sl_2)$-module
with respect to the action of the operators $\Delta(S^\pm)$ 
and $\Delta(S^3)$. 

Eq.~(\ref{rd}) implies that $r(\lambda)$ is a function of
an operator $J$ such that 
\begin{equation}\label{J}
 J \, |j,m\rangle= j \, |j,m\rangle
\end{equation}
for any vector $|j,m\rangle$ from $D_j$. In other words,
\begin{equation}\label{rP}
 r(\lambda)= \sum_{j=0}^{2S} r_{j}(\lambda) \, {\cal P}_j \,,
\end{equation}
where ${\cal P}_j$ is the projector onto $D_j$,
i.e., ${\cal P}_k |j,m\rangle= \delta_{jk} \, |j,m\rangle $.
Taking (\ref{rP}) into account, one can solve equation 
(\ref{rll0}) explicitly (\cite{Ji1,Ji2} and see also Appendix A),
\begin{equation}\label{unr}
 r(\lambda) =  {\cal P}_0 + \sum_{j=1}^{2S} \, 
  \Bigl( \prod_{k=1}^{j} \frac{\sin[\gamma(k-i\lambda)]}%
 {\sin[\gamma(k+i\lambda)]} \Bigr) \, {\cal P}_j \,.
\end{equation}
Two obvious consequences of this formula are
\begin{equation}\label{rprop}
  r(\lambda) \,  r(-\lambda) = 1 \otimes 1 \,, \qquad
  r(\lambda) \, r(\mu) = r(\mu) \, r(\lambda) \,.
\end{equation}

Introduce a $q$-analogue of the gamma function 
satisfying the relation
\[  
 (q^x - q^{-x}) \, \Gamma_q(x) = 
 (q - q^{-1}) \, \Gamma_q(x+1)
\]
and normalized such that $\Gamma_q(1)=1$. 
If  $|q|\neq 1$, this equation can be solved in terms of
a convergent infinite product. For instance,
for $|q|<1$, the solution is given by 
\begin{equation}\label{qga2}
  \Gamma_q(x) = q^{\frac 12 x(1-x)} (q^{-1}-q)^{1-x}
 \prod_{n=0}^{\infty} \frac{1-q^{2n+2}}{1-q^{2n+2x}}  \,.
\end{equation}
In terms of the $q$-gamma function eq.~(\ref{unr}) 
can be rewritten as follows 
\begin{equation}\label{rga}
 r(\lambda)= \frac{\Gamma_q(J+1-i\lambda)}%
 {\Gamma_q(J+1+i\lambda)} \, \frac{\Gamma_q(1+i\lambda)}%
 {\Gamma_q(1-i\lambda)} 
\end{equation}
with $J$ defined by (\ref{J}). 
Eqs.~(\ref{unr}) and (\ref{rga}) are $q$-deformations 
of their $XXX$ counterparts~\footnote{
  Actually, \cite{KRS} and \cite{TTF} dealt with 
  the $XXX$ analogue of ${\bf \check{R}}(\lambda)$.
  In the $XXZ$ case, ${\bf P}$ and ${\bf \check{R}}(\lambda)$ 
  do not commute with $\Delta$ and therefore cannot be 
  represented as combinations of the projectors ${\cal P}_j$.}
found in \cite{KRS,TTF} (see also \cite{Fad}).

\section{Hamiltonian} \label{sH}
Having a solution ${\bf \check{R}}(\lambda)$ to 
equation (\ref{RRR}) such that ${\bf \check{R}}(0)={\bf P}$
(equivalently, $r(0)=1\otimes 1$), one can construct an 
integrable Hamiltonian for a closed chain in the following 
way \cite{KS,TTF,Ji2,Fad}
(the normalization is chosen for later convenience)
\begin{eqnarray}
 \nonumber 
 {\cal H} &=& {\rm i} \frac{\sin\gamma}{2\gamma} 
 \sum_{n=1}^N  {\bf P}_{n,n+1} \, 
 \frac{d}{d\lambda} {\bf \check{R}}_{n,n+1}(\lambda)
 \Bigl\vert_{\lambda=0}  \Bigr. =
 \sum_{n=1}^N {\bf P}_{n,n+1} \, H_{n,n+1} {\bf P}_{n,n+1}
 = \sum_{n=1}^N H_{n+1,n}  \,,  \\
 \label{Hq2}
 && H_{n,n+1} = {\rm i} \frac{\sin\gamma}{2\gamma} \frac{d}{d\lambda} 
 \ln r_{n,n+1}(\lambda) \Bigl\vert_{\lambda=0}  \Bigr.
\end{eqnarray}
with $H_{N,N+1} \equiv H_{N,1}$ (the periodic boundary
conditions). Although $H_{n,n+1} \neq H_{n+1,n}$, we will
show below that for the total Hamiltonian we have
\begin{equation}\label{Hq3}
 {\cal H} = \sum_{n=1}^N H_{n+1,n} = 
 \sum_{n=1}^N H_{n,n+1} \,.
\end{equation}

The Hamiltonian ${\cal H}$ commutes with the higher 
quantum integrals of motion which are constructed as 
higher derivatives of~$\ln r(\lambda)$. Moreover, if 
the reference state  $\omega$ for an \hbox{$L$-operator} 
in question is such that $\omega \otimes \omega$
is an eigenvector of $r(\lambda)$, then the corresponding
Bethe vectors are eigenvectors of the Hamiltonian 
\cite{BA,KS,TTF,Fad}. In particular, this is the case 
for the $L$-operator~(\ref{LS}), for which the reference 
state $\omega$ is just the highest weight vector.

Combining (\ref{rga}) with  (\ref{Hq2}), we obtain a
compact formula for the Hamiltonian,
\begin{equation}\label{HJ2}
 H_{n,n+1} = \frac{\sin\gamma}{\gamma} \,
 \bigl( \Psi_q(J_{n,n+1}+1) - \Psi_q(1) \bigr) \,,
\end{equation}
where $\Psi_q(x)$ stands for the logarithmic derivative
of $\Gamma_q(x)$. 

In order to find the Hamiltonian explicitly in terms 
of the spin generators we first substitute eq.~(\ref{unr}) 
into eq.~(\ref{Hq2}) and derive 
\begin{equation}\label{HJ}
 H_{n,n+1} =
 (\sin\gamma) \, \sum_{j=1}^{2S} \, \Bigl( \sum_{k=1}^{j}
 \frac{\cos \gamma k}{\sin \gamma k} \Bigr) \, {\cal P}_j \,.
\end{equation}
Next, we have to construct the projectors ${\cal P}_j$ 
explicitly. Recall that the Casimir operator of 
the algebra (\ref{cS}) is given by
\begin{equation}\label{C}
 C = S^- S^+ + \frac{\sin\gamma S^3 \,
 \sin\gamma(S^3+1)}{\sin^2\gamma} \,.
\end{equation}
Its value in the highest weight representation of 
spin $S$ is 
\[  
 C_S = \frac{\sin\gamma S \, \sin\gamma (S+1)}{\sin^2\gamma} =
 \frac{\cos\gamma - \cos\gamma(2S+1)}{2\sin^2\gamma} \,. 
\]
Applying the co-multiplication (\ref{del}) to the Casimir 
operator (\ref{C}), we obtain an operator that acts in 
the tensor product $D_S \otimes D_S$,
\begin{eqnarray}\label{dC2}
 & {\bf X}_{S} = \frac 12 \Delta C 
 =  \frac 12 (q^{S^3} \, S^+) \otimes 
 (S^- \, q^{-S^3}) + \frac 12 (q^{S^3} \, S^-) 
 \otimes (S^+ \, q^{-S^3})  & \\
 & + \nonumber
 \frac{1}{4\sin^2\gamma} \Bigl( 
 (1\otimes 1 + q^{2 S^3} \otimes 
 q^{-2 S^3})\, \cos\gamma -  
 (1 \otimes q^{-2 S^3} + q^{2 S^3} 
 \otimes 1)\, \cos\bigl(\gamma(2S+1)\bigr) \Bigr) \,. &
\end{eqnarray}
Here we employed the commutation relations (\ref{cS}) 
and used (as reflected in the subscript of~${\bf X}_{S}$)
that on $D_S\otimes D_S$ we have 
$C\otimes 1=1\otimes C = (1\otimes 1) C_S$.
It is now easy to see that 
\begin{equation}\label{dCj}
 {\bf X}_{S}  =
 \frac{\sin\gamma J \ \sin\gamma (J+1)}{2\sin^2\gamma}  \,
\end{equation}
with $J$ defined by (\ref{J}). Indeed,
by construction, ${\bf X}_{S}$ commutes with $\Delta(S^\pm)$.
Bearing this in mind, (\ref{dCj}) follows from computing 
the action of ${\bf X}_{S}$ on highest weight vectors. 

For a generic $q$, the eigenvalues of (\ref{dCj}) on different 
subspaces $D_j$ in (\ref{tpr}) do not coincide. Therefore, 
we can utilize (\ref{dCj}) to construct the projectors 
${\cal P}_j$ by means of the Lagrange interpolation,
\begin{equation}\label{Pj}
 {\cal P}_j  = \prod_{ {l=0}\atop{l\neq j} }^{2S} \, 
 \frac{ 2{\bf X}_{S} -[l]_q [l+1]_q}{ [j-l]_q [j+l+1]_q} \,.
\end{equation}
Here we used the $q$-numbers defined as 
$[k]_q \equiv (q^k-q^{-k})/(q-q^{-1})=
 \frac {\sin(\gamma k)}{\sin\gamma}$. Note that the
denominator in (\ref{Pj}) can be written differently
with the help of the identity
$[j-l]_q [j+l+1]_q = [j]_q [j+1]_q - [l]_q [l+1]_q$.
We have chosen the first expression to indicate singularities
that can occur if $q$ is a root of unity.

Combining (\ref{Pj}) with (\ref{HJ}), we obtain 
a local~\footnote{
   We call $H_{n,n+1}$ local because it is a lattice 
   analogue of the Hamiltonian density. But $H_{n,n+1}$
   is also local in the sense that it involves spines
   only at two nearest sites of the lattice. We hope that
   this mixed terminology will not lead to a confusion.} 
integrable Hamiltonian for the $XXZ$ spin chain:
\begin{equation}\label{HJJ}
 H_{n,n+1} =
 (\sin\gamma) \, \sum_{j=1}^{2S} \, \biggl[ \Bigl( \sum_{k=1}^{j}
 \frac{\cos \gamma k}{\sin \gamma k} \Bigr) \, 
  \prod_{ {l=0}\atop{l\neq j} }^{2S} \, 
 \frac{ 2 (\sin\gamma)^2 \, {\bf X}_{S}  - 
 \sin \gamma l \, \sin \gamma (l+1) }{
 \sin \gamma (j-l) \, \sin \gamma (j+l+1) } \, \biggr] 
\end{equation}
with ${\bf X}_{S} \in \mathfrak{H}_n \otimes \mathfrak{H}_{n+1}$
given by (\ref{dC2}) which can also be rewritten as follows 
\begin{eqnarray}\label{Hnn}
 & {\bf X}_{S}  =  e^{{\rm i}\gamma S^3_n} \, \Bigl(
 \frac 12 S^+_n S^-_{n+1} + \frac 12 S^-_n S^+_{n+1} \Bigr. 
 \nonumber   
 + \sin\gamma S^3_n \sin\gamma S^3_{n+1} 
 \frac{\cos\gamma S \cos\gamma (S+1)}{\sin^2\gamma} & \\ [1mm]
 &  + \cos\gamma S^3_n \cos\gamma S^3_{n+1} 
 \frac{\sin\gamma S \sin\gamma (S+1)}{\sin^2\gamma} 
 \Bigl. \Bigr) \, e^{- {\rm i}\gamma S^3_{n+1}}  \,. &
\end{eqnarray}
We remark that $2{\bf X}_{S}$ is a $q$-deformation 
of square of the sum of two spins in the sense 
that in the $\gamma\rightarrow 0$ limit 
eq.~(\ref{Hnn}) simplifies to
\[  
 {\bf X}_{S}^0 = S(S+1) + {\bf S}_n^0 \cdot {\bf S}_{n+1}^0 
 = \frac 12  ({\bf S}_n^0 + {\bf S}_{n+1}^0)^2\,,
\]
where $S_a^0$ are the generators of $sl_2$. In this 
limit eq.~(\ref{HJJ}) turns into the integrable 
$XXX$ Hamiltonian constructed in \cite{Bab,TTF}.

Let us consider the large spin $S$ asymptotics of the 
Hamiltonian (\ref{HJ2})-(\ref{HJ}) in terms of the
spin operator $J$ defined by~(\ref{J}). 
Recall first that for the logarithmic derivative of 
the (non-deformed) gamma function we have
\[
 \Psi(j+1) = \Psi(1)  +
 \sum_{k=1}^j \frac 1k = \ln (j+ \frac 12) + O(j^{-2}) \,.
\]
Therefore the large spin approximation of the $XXX$
Hamiltonian is
\begin{equation}\label{Hx}
 H_{n,n+1} = \ln (J+ \frac 12) + {\rm const} \,.
\end{equation}
In the $XXZ$ case, if $q$ is real, eq.~(\ref{qga2}) 
(or its counterpart for $|q|>1$) yields
$\Psi_q(x) \approx x \, |\gamma |$ for large $x$.
More precisely, we infer from (\ref{HJ2})-(\ref{HJ}) that
\begin{equation}\label{Psiz}
 \Psi_q(j+1) = \Psi_q(1)  +
 |\gamma| \sum_{k=1}^j \tanh(k |\gamma| ) = 
 |\gamma| \, j+ \varkappa_\gamma + O(e^{-2|\gamma|j}) \,,
\end{equation}
where $\varkappa_\gamma$ is a $\gamma$-dependent constant.
Hence the large spin asymptotics of the $XXZ$ Hamiltonian is
\begin{equation}\label{Hz}
 H_{n,n+1} = J \, \sinh |\gamma| + {\rm const} \,.
\end{equation}
Here we should remind that the spin operator in (\ref{Hz})
is not the same as in (\ref{Hx}), Indeed, it is always given 
by $J= \sum_j j {\cal P}_j$ but the projectors are different
for different~$\gamma$. Notice also that the correction 
to the leading order in (\ref{Psiz}) decays very fast 
(because $\tanh(x) = 1 + O(e^{-2x})$).
So, if $|\gamma|S$ is not too small, (\ref{Hz}) is a good 
approximation even for not too large~$S$.

Thus, the large spin asymptotics of $H_{n,n+1}$ in the $XXZ$ 
case differs from that in the $XXX$ case. Moreover, for real 
$\gamma$, i.e., when $|q|=1$, the function $\Psi_q(x)$ is 
not monotonous (as seen from (\ref{HJ})) and does not have 
an asymptotics at all.

\section{Properties}\label{pR}
{\em Global symmetry:}\ 
Recall that the co-associativity property,
$(\Delta \otimes 1) \Delta = (1 \otimes  \Delta) \Delta$,
leads to a natural notion of an $n$-th power of 
the co-multiplication: 
$\Delta^{(n+1)}\equiv (\Delta \otimes 1^{\otimes n})
 \Delta^{(n)}$ with $\Delta^{(1)}\equiv \Delta$. 
Then the global spin generators for a chain with $N$ nodes
are naturally introduced~\cite{Ji1} as 
${\cal S}^\pm = \Delta^{(N-1)} (S^\pm)$ and 
${\cal S}^3 = \Delta^{(N-1)}(S^3)$; 
which in explicit form reads
\begin{equation}\label{tot}
 {\cal S}^3 = \sum_{n=1}^N S^3_n \,, \qquad
 {\cal S}^\pm =  
 \sum_{n=1}^N q^{S^3_1} \ldots q^{S^3_{n-1}} 
 S^\pm_n q^{-S^3_{n+1}} \ldots q^{-S^3_N} \,.
\end{equation}
By construction (see (\ref{rd})
and (\ref{HJ})), the local Hamiltonian $H_{n,n+1}$ is 
$U_q(sl_2)$-symmetric, i.e., it commutes with 
$\Delta(\xi)$ for any $\xi\in U_q(sl_2)$. Furthermore,
it is easy to check that
\begin{eqnarray}
 \label{symm3}
 && [\, H_{n,n+1} , {\cal S}^3 \,] = 0 \,, \qquad
 {\rm for}\ n=1,\ldots,N \,, \\
 \label{symmS}
 && [\, H_{n,n+1}, {\cal S}^\pm \,] = 0 \,, \qquad 
 {\rm for}\ n=1,\ldots,N-1 \,.
\end{eqnarray}
Since $H_{N,1}$ does not satisfy (\ref{symmS}), the total 
Hamiltonian enjoys only the $U(1)$-symmetry,
\begin{equation}\label{HS3}
  [\, {\cal H}, {\cal S}^3 \,] = 0 \,.
\end{equation}
Actually, the higher quantum integrals of motion also 
commute with~${\cal S}^3$. 
Therefore in the presence of a constant magnetic 
field~$h$ the corresponding Hamiltonian,
${\cal H}_h = {\cal H} + h {\cal S}^3$, remains 
integrable and $U(1)$-symmetric.

{\em $C$ and $P$ symmetries:}\ Recall that ${\bf P}$ denotes 
the permutation in $\mathfrak{H}\otimes \mathfrak{H}$. 
Since the co-multiplication $\Delta$ does not commute with 
$\bf P$, neither does $r(\lambda)$, ${\bf X}_S$ or $H_{n,n+1}$. 
However, we observe that (here and below dependence on 
$\gamma$ is shown explicitly only if it is affected by a
transformation)
\begin{equation}\label{rPg}
 {\bf P} \, {\bf X}_S(\gamma) \, {\bf P} = {\bf X}_S(-\gamma)
 \,, \qquad
 {\bf P} \, r(\lambda,\gamma) \,{\bf P} = r(\lambda,-\gamma) \,.
\end{equation}
The first relation is obvious from (\ref{Hnn}) and yields
the second one upon noticing that both the projectors 
${\cal P}_j(x,\gamma)$ and the coefficients 
$r_j(\lambda,\gamma)$ in (\ref{rP}) are even functions 
in $\gamma$. Further, the second relation 
in (\ref{rPg}) implies readily that
\begin{equation}\label{Pg}
 {\bf P}_{n,n+1} \, H_{n,n+1}(\gamma) \, {\bf P}_{n,n+1} 
 = H_{n,n+1}(-\gamma) \,.
\end{equation}
Thus $H_{n,n+1}$ does not have the $P$ (reflection) 
symmetry. But, if $q$ is real, it has the \hbox{$C$-symmetry}
(invariance with respect to the complex conjugation).
If $|q|=1$, then eq.~(\ref{Pg}) shows that 
$H_{n,n+1}$ has no $C$- or $P$-symmetry
separately but it has the $CP$-symmetry.  

{\em Local bulk and boundary terms:}\ 
As we will see below, $H_{n,n+1}$ decomposes into two 
parts (which we will refer to as the local
bulk term and the local boundary term):
\begin{equation}\label{dec}
  H_{n,n+1} = \widehat{H}_{n,n+1} +
 {\rm i} \frac{\sin\gamma}{2} (S^3_n - S^3_{n+1}) \,.
\end{equation}
The local bulk term, $\widehat{H}_{n,n+1}$, has
the following properties (see Section~\ref{tL})
\begin{equation}\label{Hh}
{\bf P}_{n,n+1} \widehat{H}_{n,n+1} {\bf P}_{n,n+1}
 = \widehat{H}_{n,n+1} \,, \qquad
\widehat{H}_{n,n+1}(\gamma)=\widehat{H}_{n,n+1}(-\gamma) \,.
\end{equation}
Thus, $\widehat{H}_{n,n+1}$ is $P$- and $C$-symmetric
for real $q$ as well as for $|q|=1$.
In fact (see Section~\ref{tL}),
$\widehat{H}_{n,n+1}$ is a local Hamiltonian
associated with the $L$-operator
\begin{equation}\label{Lh}
  \widehat{L}(\lambda) = \frac{1}{\sin\gamma} \left( 
 \begin{array}{cc} \sinh [\gamma(\lambda + {\rm i} S^3)] &
  {\rm i}\, S^- \sin\gamma  \\ 
 {\rm i}\, S^+ \sin\gamma &
 \sinh [\gamma(\lambda - {\rm i} S^3)] 
 \end{array} \right) \,.
\end{equation}
Notice that $\widehat{H}_{n,n+1}$ is not $U_q(sl_2)$-symmetric
but only $U(1)$-symmetric.

The local boundary term in (\ref{dec}) has the same form for 
all positive (half-) integer spins. In the total Hamiltonian 
of a closed chain all these boundary terms mutually cancel,   
hence
\begin{equation}\label{HHh}
 {\cal H} = \sum_{n=1}^N H_{n,n+1} = 
 \sum_{n=1}^N \widehat{H}_{n,n+1} \,.
\end{equation}
This together with (\ref{Hh}) explains why the 
two sums in (\ref{Hq3}) coincide.

{\em Properties with respect to $*$--operation:}\
The $*$--structure on $U_q(sl_2)$ 
corresponding to the compact real form $U_q(su(2))$
is defined as an anti-automorphism such that
\begin{equation}\label{star}
  (S^\pm)^* = \eta^{\pm 1} \, S^\mp \,, \qquad  (S^3)^* = S^3 
\end{equation}
with some real $\eta$. Of course, the prefactor 
$\eta^{\pm 1}$ can be eliminated by rescaling the generators.
However it may be convenient to keep it. For instance, we 
saw in Section~\ref{uR} that, in the spin-$\frac 12$ case,
it is natural to put $S^\pm =q^{\pm \frac 12} \sigma^\pm$. 
Then $\eta=1$ if $|q|=1$ but $\eta=q$ if $q$ is real.
In fact, the choice of $\eta$ is not important 
for our purposes since $H_{n,n+1}$ and $\widehat{H}_{n,n+1}$
contain $S^\pm$ only in homogeneous combinations 
like $(S^+ \otimes S^-)$.

The action of the $*$--operation extends on a tensor 
product as $(\xi\otimes \zeta)^* = \xi^* \otimes \zeta^*$. 
It must be remarked that properties of the co-multiplication
with respect to the action of the $*$--operation depend 
on the choice of $q$, namely we infer from (\ref{del})
and (\ref{star}) that~\footnote{
  A \hbox{$*$--structure} 
  satisfying $(\Delta(\xi))^* = \Delta(\xi^*)$
  for $|q|=1$ is the {\em non-compact} real form 
  $U_q(sl(2,{\mathbb R}))$, see \cite{su}.} 
\[
 (\Delta(\xi))^* = \Delta(\xi^*) \quad
  {\rm if}\ q \in  {\mathbb R} \qquad {\rm but} \qquad
 (\Delta(\xi))^* = {\bf P} \Delta(\xi^*) {\bf P} \quad
  {\rm if}\ |q|=1 \,. 
\]
Therefore the local Hamiltonian (\ref{HJJ}) has the 
following properties:
\begin{eqnarray}
 \label{*1}
 && (H_{n,n+1})^* = H_{n,n+1} \qquad
  {\rm for}\ q \in  {\mathbb R} \\
 \label{*2} 
 && (H_{n,n+1})^* = 
 {\bf P}_{n,n+1} H_{n,n+1} {\bf P}_{n,n+1}  \qquad
  {\rm for}\  |q|=1 \,. 
\end{eqnarray}
Nevertheless, we see from eq.~(\ref{HJ}) that the 
eigenvalues of $H_{n,n+1}$ are real in the both cases.
Furthermore, it follows from (\ref{dec})-(\ref{Hh}) 
and (\ref{*1})-(\ref{*2}) that
\begin{equation}\label{Hh*}
 ({\widehat{H}}_{n,n+1})^* = \widehat{H}_{n,n+1} 
\end{equation}
for the both regimes of $q$. In combination 
with (\ref{HHh}) this implies that the relation
\begin{equation}\label{H*}
 {\cal H}^* = {\cal H} 
\end{equation}
holds also in the both regimes of $q$. 

It is worth emphasizing that, for $|q|=1$, objects which 
have the global $U_q(sl_2)$-symmetry, like $H_{n,n+1}$ 
for $n \neq N$, are in general not self-conjugate with 
respect to the $*$--operation~(\ref{star}). Indeed, if 
${\cal O}^*={\cal O}$ and it commutes with the 
global spin generators ${\cal S}^\pm$ given by (\ref{tot}),
then it must also commute with $({\cal S}^\pm)^*$. This
imposes a strong extra condition on the structure 
of $\cal O$ since for $|q|=1$ the conjugate of ${\cal S}^\pm$ 
belong to $U_{q^{-1}}(sl_2)$ rather than to $U_q(sl_2)$.
Vice versa, objects that are self-conjugate for $|q|=1$
are in general not  $U_q(sl_2)$-symmetric, for
instance $\widehat{H}_{n,n+1}$ and ${\cal H}$.

{\em $T$ symmetry:}\
The definition $(\xi\otimes \zeta)^* = \xi^* \otimes \zeta^*$
is consistent with the property of the matrix transposition,
$(\xi \otimes \zeta)^t = \xi^t \otimes \zeta^t$, if 
$\xi$ and $\zeta$ are regarded as matrices. Thus 
the $*$-operation can be realized as the matrix hermitian 
conjugation, $\xi^*=\bar{\xi}{\,}^t$ (bar denotes the 
complex conjugation).

 {}From (\ref{*1}), (\ref{*2}) and (\ref{Pg}) we deduce
that $(H_{n,n+1})^*=\bar{H}_{n,n+1}$ in the both regimes
of~$q$. Therefore finite dimensional matrix representations 
of the considered above Hamiltonians are symmetric matrices,
\begin{equation}
 \label{Ht} 
 (H_{n,n+1})^t = H_{n,n+1} \,, \qquad
 (\widehat{H}_{n,n+1})^t = \widehat{H}_{n,n+1} \,, \qquad
 ({\cal H})^t = {\cal H} \,.
\end{equation}
Here the first equality leads to the second and the third
due to (\ref{dec}) and (\ref{HHh}), respectively.

\section{Examples ($S= \frac 12 , 1 , \frac 32$)}
\label{eX}
For the spin $S=\frac 12$, the Hamiltonian (\ref{HJJ}) 
is very simple:
\begin{eqnarray}\label{pr1}
 & H_{n,n+1} = (\cos\gamma) \, {\cal P}_1 =  
 {\bf X}_{\frac 12} \,, &
\end{eqnarray}
and the spin generators are given by
$S^\pm = q^{\pm \frac 12} \sigma^\pm$ and 
$S^3 = \frac 12 \sigma^3$. Using the relations
\begin{equation}\label{ex1}
 e^{t \sigma^3} \sigma^\pm = \sigma^\pm 
 e^{-t \sigma^3} = e^{\pm t } \sigma^\pm
 \,, \qquad  e^{{\rm i} t \sigma^3} = 
 \cos t + {\rm i} \sigma^3 \, \sin t  \,,
\end{equation}
it is easy to check that (\ref{pr1}) acquires the following form
\begin{eqnarray}\label{H12}
  H_{n,n+1} = \frac 12 (S^+_n S^-_{n+1} + S^-_n S^+_{n+1})
 + (\cos\gamma) \, S^3_n S^3_{n+1} + \frac 34 \cos\gamma
 + {\rm i} \frac{\sin\gamma}{2} (S^3_n - S^3_{n+1}) \,.
\end{eqnarray}
The bulk term here is the well-known $XXZ$ deformation 
of the Heisenberg spin chain. 

For $S=1$ the Hamiltonian (\ref{HJJ}) looks as follows
\begin{equation}
 \label{H1a} 
 H_{n,n+1} = (\cos\gamma) \, {\cal P}_1
 + \frac{\sin 3\gamma}{\sin 2\gamma}  \, {\cal P}_2 
 = \frac{1}{4 (\cos\gamma)^3 } 
 \Bigl( (\cos\gamma + 4 \cos^3\gamma) \, {\bf X}_{1} - 
  ({\bf X}_{1})^2 \Bigr) \,. 
\end{equation}
In this case the spin generators are given by (\ref{S1z}); 
they are related to the spin-1 \hbox{$sl_2$-generators} as 
$S^\pm = (\cos\gamma)^{\frac 12} S^\pm_0$, $S^3=S^3_0$.
Rewriting ${\bf X}_{1}$ and its square as polynomials 
in the spin generators (see Appendix B), we obtain
\begin{equation}\label{H1Sb}
 H_{n,n+1} = \frac{1}{4\cos\gamma} 
 \Bigl(  {\bf \Omega}_1 - ({\bf \Omega}_1)^2 
 + {\bf F}_1 \Bigr) + {\rm i} \frac{\sin\gamma}{2} 
 ( S^3_{n} - S^3_{n+1} )  \,,
\end{equation}
where 
\begin{eqnarray} 
 \label{om1}
 {\bf \Omega}_1 &=& \frac{1}{2\cos\gamma} \bigl(S^+_{n} S^-_{n+1}
 + S^-_{n} S^+_{n+1} \bigr) + 
 (2\cos\gamma-1) \, S^3_{n} S^3_{n+1} \,, \\
 \label{F1}
 {\bf F}_1 &=& 2 \cos\gamma (\cos\gamma-1) S^3_{n} S^3_{n+1} 
  + 2 (\cos\gamma-1)^2  (S^3_{n} S^3_{n+1})^2 \\ [0.5mm]
 \nonumber
 &&  -\, 2 (\sin\gamma)^2 \, \bigl(
 (S^3_{n})^2 + (S^3_{n+1})^2 \bigr) + 4 + 2\cos 2\gamma \,.
\end{eqnarray}
The bulk term of the Hamiltonian (\ref{H1Sb}) (i.e.,
(\ref{H1Sb}) without the last term) coincides (up to a constant) 
with the Fateev-Zamolodchikov (FZ) Hamiltonian \cite{FZ}.
The FZ~Hamiltonian is usually written in a slightly different
way in terms of the spin-1 $sl_2$-generators; we have chosen
the above form since it allows for better comparison with 
the $S=\frac 32$ case.

In the limit $\gamma\rightarrow 0$, (\ref{H1a})-(\ref{H1Sb}) 
simplifies to the well-know spin-1 $XXX$ Hamiltonian:
\begin{eqnarray*}
 & H_{n,n+1}^0 = \frac{5}{4} \, {\bf X}_1^0 - \frac{1}{4} \, 
  ({\bf X}_1^0 )^2 = \frac{1}{4} \,
 {\bf S}_n^0 \cdot {\bf S}_{n+1}^0 - \frac{1}{4} \,
 ({\bf S}_n^0 \cdot {\bf S}_{n+1}^0 )^2  + \frac{3}{2} \,.  &
\end{eqnarray*}

In the $S=\frac 32$ case the Hamiltonian (\ref{HJ}) is given by
\begin{equation} \label{H32a} 
 H_{n,n+1} =  (\cos\gamma)\, {\cal P}_1
 + \frac{\sin 3\gamma}{\sin 2\gamma} \, {\cal P}_2 +
 \Bigl( \frac{\sin 3\gamma}{\sin 2\gamma} + (\sin\gamma)
 \frac{\cos 3\gamma}{\sin 3\gamma} \Bigr) \, {\cal P}_3 \,.
\end{equation}
Rewriting the corresponding expression (\ref{HJJ}) in 
the polynomial (with respect to the spin generators) form
(see Appendix B), we obtain 
\begin{eqnarray}
 \nonumber
 H_{n,n+1} &=& \frac{1}{12(\cos\gamma)^3 (1+2\cos 2\gamma)^3} 
 \Bigl( 12(\cos\gamma)\, ({\bf \Omega}_{\frac 32})^3 + 
 (5\cos 4\gamma- \cos 2\gamma +2)\, 
 ({\bf \Omega}_{\frac 32})^2  \\
 \label{H32Sb}
 && + \, \frac{1}{2\cos\gamma} 
 \bigl( {\bf \Omega}_{\frac 32} \, {\bf Q} 
 + {\bf Q} \, {\bf \Omega}_{\frac 32}  \bigr)
 + \frac{1+2\cos 2\gamma}{64 \cos^2\gamma} {\bf F}_{\frac 32}  
 \Bigl. \Bigr)  + {\rm i} \frac{\sin\gamma}{2} 
 ( S^3_{n} - S^3_{n+1} )   \,, 
\end{eqnarray}
where ${\bf Q}$ and ${\bf F}_{\frac 32}$ are real symmetric 
polynomials in $S^3_{n}$ and $S^3_{n+1}$ (see Appendix B) and
\begin{equation} \label{om32}
  {\bf \Omega}_{\frac 32} = \frac 12 
 \bigl(S^+_{n} S^-_{n+1} + S^-_{n} S^+_{n+1} \bigr) + 
 (1+2\cos 2\gamma) \frac{\cos 2\gamma}{3\cos\gamma} \,
 S^3_{n} S^3_{n+1} \,.
\end{equation}

Let us remark that in both Hamiltonians (\ref{H1Sb})
and (\ref{H32Sb}) the two ``most non-diagonal'' terms 
are powers of an operator ${\bf \Omega}$ which is
quadratic in generators. To clarify this fact, we observe
that, like for $S= \frac 12$, the corresponding
Casimir operators (\ref{C}) are actually quadratic: 
\begin{eqnarray}
 \label{Cn1}
 && \frac 12 (S^+ S^- + S^- S^+) + (\cos\gamma) (S^3)^2 =
 2\cos\gamma  \qquad {\rm for}\ S=1 \,, \\
 \label{Cn32}
 && \frac 12 (S^+ S^- + S^- S^+) + (\cos\gamma)^2 (S^3)^2 =
 \frac 14 (17 \cos^2\gamma -2 ) \qquad {\rm for}\ S=\frac 32 
\end{eqnarray}
(the last expression differs from $C_{\frac 32}$ by a 
constant). Eq.~(\ref{Cn1}) is due to the first
relation in~(\ref{s1a}); eq.~(\ref{Cn32}) follows 
 {}from the relation (\ref{ex3b}) and the identity
$2(\sin \gamma S^3)^2 = 1 - \cos 2\gamma S^3$.

\section{Another co-multiplication}
\label{tC}
As we saw in Section~\ref{uR}, the co-multiplication 
operation $\Delta$ plays a key r{\^o}le in the 
construction of the Hamiltonian $\cal H$. However, 
eq.~(\ref{del}) does not exhaust possible definitions 
of $\Delta$ for $U_q(sl_2)$. In fact, a homomorphism 
$\Delta_\alpha : U_q(sl_2) \rightarrow 
 [U_q(sl_2)]^{\otimes 2}$ such that 
\begin{equation}\label{dela}
 \Delta_\alpha(S^\pm) = S^\pm \otimes q^{ (\pm\alpha -1) S^3} 
 + q^{(\pm\alpha +1) S^3} \otimes S^\pm \,,
 \qquad \Delta_\alpha(S^3) = S^3 \otimes 1 + 1 \otimes S^3 
\end{equation}
satisfies all the properties of the co-multiplication
for any real $\alpha$. 
We had before $\alpha =0$ which is the most ``symmetric'' 
choice. But we may prefer $\alpha= 1$ (or $\alpha= -1$) if 
we want the regime $q<0$ to be on equal footing with $q>0$ 
in the sense that the terms $q^{(\pm\alpha \pm 1) S^3}$  
in~(\ref{dela}) take only real values.

Notice that $\Delta_\alpha$ is obtained from the
co-multiplication (\ref{del}) by twisting:
\begin{equation}\label{twi1}
 \Delta_\alpha(\xi) = F_\alpha \,
 \Delta(\xi) \, (F_\alpha)^{-1} \,, \qquad
 F_\alpha = q^{\alpha S^3 \otimes S^3} 
 \in \mathfrak{H} \otimes \mathfrak{H} \,.
\end{equation}
This twist is rather specific in that it preserves 
the co-associativity of the co-multiplication (more general 
twists give rise to the so-called quasi-Hopf algebras 
\cite{twi1}; see \cite{twi2} for their 
applications in integrable spin models).

In order to construct an integrable Hamiltonian using
the new co-multiplication~$\Delta_\alpha$ in the same 
way as we used~$\Delta$ in Sections~\ref{uR} and~\ref{sH}, 
we should first find an $L$-operator,
$\tilde{L}_\alpha (\lambda)$, that satisfies~(\ref{Ldel}) 
with~$\Delta_\alpha$. For this purpose we observe that
\begin{equation}\label{fF}
 F_\alpha = (\phi_\alpha \otimes \phi_\alpha)^{-1} \,
 \Delta(\phi_\alpha) \qquad  {\rm where}\quad 
 \phi_\alpha = q^{\frac{\alpha}{2} (S^3)^2} \in \mathfrak{H} \,.
\end{equation}
Therefore we can rewrite the l.h.s.~of (\ref{Ldel}) 
for $\tilde{L}_\alpha (\lambda)$ as
\[
 \Delta_\alpha ( \tilde{L}_\alpha (\lambda) ) =
 (\phi_\alpha \otimes \phi_\alpha)^{-1} \,
 \Delta \bigl( \phi_\alpha \, \tilde{L}_\alpha (\lambda) \, 
 \phi_\alpha^{-1} \bigr) \, (\phi_\alpha \otimes \phi_\alpha) 
\]
which makes it obvious that $\tilde{L}_\alpha (\lambda)$
satisfies~(\ref{Ldel}) with~$\Delta_\alpha$ if it
is related to the $L$-operator (\ref{LS}) as follows
\begin{eqnarray}
 \label{LSa}
 \tilde{L}_\alpha (\lambda) &=& 
 \phi_\alpha^{-1} \, L(\lambda) \, \phi_\alpha  \\ [0.5mm]
 \label{LSb}
 &=&  \frac{1}{\sin\gamma} \left( 
 \begin{array}{cc} \sinh [\gamma(\lambda + {\rm i} S^3)] &
  {\rm i}\, \sin\gamma \, e^{\gamma\lambda} 
 q^{ \frac{\alpha}{2}} q^{ \alpha S^3} \, S^-   \\  
 {\rm i}\, \sin\gamma \, e^{-\gamma\lambda} 
 q^{ \frac{\alpha}{2}} q^{- \alpha S^3} \, S^+  &
 \sinh [\gamma(\lambda - {\rm i} S^3)] 
 \end{array} \right) .
\end{eqnarray}
Since the map 
$\xi \rightarrow \phi_\alpha^{-1} \xi \phi_\alpha$
is an automorphism of $U_q(sl_2)$, it is clear that
the exchange relation (\ref{RLL}) holds for 
$\tilde{L}_\alpha (\lambda)$ with the same 
\hbox{$R$-matrix} as for~$L(\lambda)$. Consequently, 
the Bethe ansatz equations for $\tilde{L}_\alpha (\lambda)$
coincide with those for $L(\lambda)$ 
(the reference state $\omega$ has also not changed).

Substituting (\ref{LSa}) into (\ref{rLL}),
we find a universal $r$-matrix for the new
\hbox{$L$-operator}:
\begin{equation}\label{rf}
 \tilde{r}_\alpha(\lambda) =  
  (\phi_\alpha \otimes \phi_\alpha)^{-1} \,
  r(\lambda) \, (\phi_\alpha \otimes \phi_\alpha) 
  = F_\alpha \, r(\lambda) \, (F_\alpha)^{-1} \,.
\end{equation}
The second equality is due to relation (\ref{fF}) 
and the property (\ref{rd}).
The same twists relate the corresponding solutions of 
the Yang-Baxter equation (\ref{RRR}), i.e.,
\begin{equation}\label{Rf}
{\bf \check{R}}_\alpha(\lambda) =
 (\phi_\alpha \otimes \phi_\alpha)^{-1}
 {\bf \check{R}}(\lambda) (\phi_\alpha \otimes \phi_\alpha)
 = F_\alpha \,{\bf \check{R}}(\lambda) (F_\alpha)^{-1} \,.
\end{equation}
Notice that the first equality is consistent with (\ref{LSa})
since $\phi_\alpha$ in the fundamental representation is 
just a constant. Further, evaluating the r.h.s.~of 
(\ref{Rf}) in $V \otimes \mathfrak{H}$ representation, we 
see that $\tilde{L}_\alpha(\lambda)$ can be constructed 
also as a twist by 2$\times$2 matrix,
\[
 \label{LSaa}
 \tilde{L}_\alpha (\lambda) = 
 f_\alpha \, L(\lambda) \, f_\alpha^{-1} \,, \qquad
 f_\alpha = q^{\frac{\alpha}{2} \sigma^3 \otimes S^3 } \,.
\]
Let us underline that existence of the two ways 
of constructing $\tilde{L}_\alpha(\lambda)$ and, 
as a consequence, of the relation
\[
 [ L(\lambda),  \phi_\alpha \, f_\alpha ] = 0 
\]
is due to the property of the universal $r$-matrix 
(\ref{rd}) applied to $\xi= \phi_\alpha$ (indeed,
$q^{\frac{\alpha}{2}}\phi_\alpha f_\alpha$ is 
$\Delta(\phi_\alpha)$
evaluated in $V \otimes {\mathfrak H}$).

According to (\ref{Hq2}), the local Hamiltonian 
corresponding to the universal $r$-matrix (\ref{rf})~is
\begin{equation}\label{Ha}
 \tilde{H}_{n,n+1}^{(\alpha)} =
  (\phi^\alpha_n \, \phi^\alpha_{n+1})^{-1} \,
  H_{n,n+1} \, (\phi^\alpha_n \, \phi^\alpha_{n+1})
  = F^\alpha_{n,n+1} \,
  H_{n,n+1} \, (F^\alpha_{n,n+1})^{-1} \,.
\end{equation}
This transformation does not modify the $S=\frac 12$ 
Hamiltonian (\ref{pr1}) since in this case $\phi_\alpha$
is trivial. But already for $S=1$ we find (see Appendix B)
\begin{equation}\label{H1Sc}
 \tilde{H}_{n,n+1}^{(\alpha)} - H_{n,n+1} =
 \frac{1-\cos(\alpha\gamma)}{2\cos\gamma} \,
 \bigl\{ \Upsilon_{n,n+1} , S^3_{n} S^3_{n+1} \bigr\} 
 - {\rm i} \frac{\sin(\alpha\gamma)}{2\cos\gamma} \,
 \bigl[ \Upsilon_{n,n+1} , S^3_{n} S^3_{n+1} \bigr] \,,
\end{equation}
where $[,]$ and $\{,\}$ stand for commutator and
anticommutator, respectively, and
$\Upsilon_{n,n+1} \equiv \frac 12
 (S^+_{n} S^-_{n+1} + S^-_{n} S^+_{n+1})$.
Notice that the terms on the r.h.s.~of (\ref{H1Sc}) 
are non-diagonal.

Finally, the total Hamiltonian is given by
\[ 
 \tilde{\cal H}^{(\alpha)} = \sum_{n=1}^{N} 
 \tilde{H}_{n,n+1}^{(\alpha)} = (\Phi_\alpha)^{-1} \,
 {\cal H} \, \Phi_\alpha =
 {\cal F}_\alpha \, {\cal H} \, ({\cal F}_\alpha)^{-1} \,, 
 \qquad \Phi_\alpha \equiv \prod_{n=1}^N \phi_\alpha \,,
 \quad   {\cal F}_\alpha \equiv 
 q^{\alpha \! \sum\limits_{n<m} \! S^3_n S^3_m} .
\]
The $\Phi$--twist here follows easily from~(\ref{Ha}).
The ${\cal F}$--twist yields the same result because 
${\cal H}$ commutes with ${\cal S}^3$ (\ref{HS3})
and hence with 
$\Phi_\alpha {\cal F}_\alpha = 
  q^{\frac{\alpha}{2} ({\cal S}^3)^2}$.
Since $\tilde{\cal H}^{(\alpha)}$ and ${\cal H}$ are
related by a twist they have the same set of eigenvalues;
this agrees with the fact that the Bethe ansatz 
equations have not changed.

Let us conclude this section with a remark:
eqs.~(\ref{rf}), (\ref{Rf}) and (\ref{Ha}) may appear 
to suggest that $\tilde{L}_\alpha(\lambda)$ and 
$\tilde{r}_\alpha(\lambda)$ are related to another 
twisted co-multiplication, 
$\tilde{\Delta}_\theta (\xi) \equiv \theta^{-1} \,
  \Delta(\xi) \, \theta$, where 
$\theta=\phi_\alpha \otimes \phi_\alpha$.
But $\tilde{\Delta}_\theta$ fails to satisfy 
the necessary property of a co-multiplication \cite{twi1}, 
$(\epsilon \otimes 1) \Delta (\xi) = \xi$,
where $\epsilon$ is the co-unit. So $\tilde{\Delta}_\theta$ 
is not a co-multiplication of a (quasi) Hopf algebra.
(Moreover, $\tilde{\Delta}_\theta$ is not 
co-associative.)

\section{From $H_{n,n+1}$ to $\widehat{H}_{n,n+1}$}
\label{tL}
Consider now another family of $L$-operators,
\begin{equation}\label{LSc}
  \widehat{L}_\beta (\lambda) = \frac{1}{\sin\gamma} \left( 
 \begin{array}{cc} \sinh [\gamma(\lambda + {\rm i} S^3)] &
  {\rm i}\, \sin\gamma \, e^{(\gamma-\beta) \lambda} \, S^- \\ 
 {\rm i}\, \sin\gamma \, e^{(\beta-\gamma) \lambda} \, S^+  &
 \sinh [\gamma(\lambda - {\rm i} S^3)] 
 \end{array} \right) 
\end{equation}
which are obtained from the $L$-operator (\ref{LS}) 
as twists by certain 2$\times$2 matrices,
\begin{equation}\label{tw3}
 \widehat{L}_\beta (\lambda) =  K_\lambda^{-1} \, 
 L(\lambda) \, K_\lambda \,, \qquad
 K_\lambda = e^{\frac 12 \beta \lambda \, \sigma^3} \in V \,.
\end{equation}
For $\beta=\gamma$ this gives the $L$-operator (\ref{Lh})
(which is most often used in applications of the Bethe ansatz). 
The map $S^\pm \rightarrow e^{\pm \beta \lambda} S^\pm$,
$S^3 \rightarrow S^3$ is an automorphism of $U_q(sl_2)$
but, unlike the case treated in Section~\ref{tC}, 
it is $\lambda$-dependent. 
Therefore, the $\check{R}$-matrix corresponding to 
$\widehat{L}_\beta (\lambda)$ differs from (\ref{R}).
Namely, as seen from (\ref{tw3}), exchange relation 
(\ref{RLL}) holds for $\widehat{L}_\beta (\lambda)$ with 
$R_\beta(\lambda)= (1 \otimes K_\lambda^{-1})
 R(\lambda) (K_\lambda \otimes 1)$, i.e.,
$R_\beta(\lambda)= P \check{R}_\beta (\lambda)$, where
\begin{equation}\label{Rb}
 \check{R}_\beta (\lambda) =  
 {\rm i}\, e^{(\gamma-\beta)\lambda} \,
 \sigma^+ \otimes \sigma^- + {\rm i}\, 
 e^{(\beta-\gamma)\lambda} \, \sigma^- \otimes \sigma^+  
   + \frac{1}{\sin\gamma}  \sinh \Bigl( \gamma\lambda +
 \frac{{\rm i}\,\gamma}{2} (1 \otimes 1 + 
   \sigma^3 \otimes \sigma^3) 
 \Bigr) \,.
\end{equation}

In order to find a universal $r$-matrix for 
$\widehat{L}_\beta (\lambda)$, we will apply the
approach which we used in Section~\ref{tC}. Namely,
we observe that 
$\varphi_\lambda \equiv e^{\beta \lambda S^3} 
 \in {\mathfrak H}$
is a complementary twist to (\ref{tw3}) in the sense that
\[
 [ L(\lambda), K_\lambda \, \varphi_\lambda  ] = 0  
\]
as follows from the property of the
universal $r$-matrix (\ref{rd}) for $\xi=\varphi_\lambda$
(in fact, $K_\lambda$ is just $\varphi_\lambda$ evaluated
in the fundamental representation; hence 
$K_\lambda \varphi_\lambda$ is $\Delta(\varphi_\lambda)$
evaluated in $V \otimes {\mathfrak H}$).

Thus, instead of the twist (\ref{tw3}) in the auxiliary 
space $V$, the $L$-operator (\ref{LSc}) can be obtained 
as a twist in the quantum space~$\mathfrak H$,
\begin{equation}\label{tw2}
 \widehat{L}_\beta (\lambda) = \varphi_\lambda \, 
 L(\lambda) \, \varphi_\lambda^{-1} \,, \qquad
  \varphi_\lambda = e^{\beta \lambda S^3} \in \mathfrak H \,.
\end{equation}

Substituting (\ref{tw2}) into (\ref{rLL}) and taking again
into account that $[r(\lambda),\Delta(\varphi_\mu)]=0$, 
we find a universal $r$-matrix for
$\widehat{L}_\beta (\lambda)$,
\begin{equation}\label{rhat}
 \widehat{r}_\beta (\lambda) =  
 (1 \otimes \varphi_\lambda) \, r(\lambda) \, 
 (\varphi_\lambda^{-1} \otimes 1) \,.
\end{equation}
Unlike eq.~(\ref{rf}) this relation is not a twist. 
Notice however that the corresponding solutions of the 
Yang-Baxter equation (\ref{RRR}) are related by a twist, 
\begin{equation}\label{Rhat}
{\bf \check{R}}_\beta(\lambda) =
 (1 \otimes \varphi_\lambda) {\bf \check{R}}(\lambda) 
 (1 \otimes \varphi_\lambda^{-1}) \,.
\end{equation}

It must be stressed now that $\widehat{L}_\beta (\lambda)$
does not possess a decomposition of the type 
\hbox{(\ref{Bor})-(\ref{Ldel})}, and 
$\widehat{r}_\beta (\lambda)$ does not commute with 
$\Delta$ (or $\Delta_\alpha$), i.e., (\ref{rd}) does not 
hold for $\widehat{r}_\beta (\lambda)$ for a generic~$\xi$.  
As a consequence, $\widehat{r}_\beta (\lambda)$ does not
have a representation of the type (\ref{rP}). Moreover,
for $\beta\neq 0$ we have in general 
$[ \widehat{r}_\beta (\lambda) , \widehat{r}_\beta (\mu)]
 \neq 0$ (except for the fundamental representation
in the case $\beta=\gamma$). Nevertheless, the general 
recipe for constructing a local integrable Hamiltonian 
applies (because the Yang-Baxter equation for 
${\bf \check{R}}_\beta(\lambda)$ is valid).
So we substitute (\ref{rhat}) into the formula 
(\ref{Hq2}) and derive
\begin{equation}
 \label{Hqhat}
 \widehat{H}_{n,n+1}^{(\beta)} = 
  H_{n,n+1} - {\rm i} \,
 \frac{\beta \, \sin\gamma}{2\gamma} \, (S^3_n - S^3_{n+1}) \,.
\end{equation}
Thus the new local Hamiltonian differs {}from 
$H_{n,n+1}$ only in the local boundary term. Hence
the total Hamiltonian
$\widehat{\cal H} = \sum_{n=1}^N \widehat{H}_{n,n+1}^{\beta}$
coincides with ${\cal H}$. 
Note that the Bethe ansatz equations, describing the spectrum 
of $\widehat{\cal H}$, have also not changed, which is not 
entirely trivial since the corresponding $R$-matrix has changed. 
The reason is that in the derivation of the Bethe ansatz 
equations the non-diagonal entries of (\ref{Rb}) appear only 
in the so-called ``unwanted terms'' that cancel each other 
\cite{KS,BA,Fad}.

In Section~\ref{pR} we asserted that the local
Hamiltonian $H_{n,n+1}$ decomposes into a local boundary
term and a local bulk term $\widehat{H}_{n,n+1}$ which
is associated with the $L$-operator (\ref{Lh}). Now this
is obvious {}from (\ref{LSc}) and (\ref{Hqhat}) if we
put $\beta=\gamma$. Since we have found the corresponding
universal $r$-matrix, we are in a position to prove
the properties of $\widehat{H}_{n,n+1}$ stated in 
Section~\ref{pR}. For brevity we denote 
$r_0(\lambda) \equiv \widehat{r}_\gamma(\lambda)$.

As seen from (\ref{Rb}) for $\beta=\gamma$, the 
auxiliary $R$-matrix associated with the $L$-operator 
(\ref{Lh}) is $P$-symmetric, i.e., it commutes with 
the 4$\times$4 permutation matrix~$P_V$. 
The corresponding universal $r$-matrix, $r_0(\lambda)$, 
has analogous properties. Namely, (as we prove in 
Appendix A) $r_0(\lambda)$ satisfies the
following relations 
\begin{eqnarray}
 \label{rg}
 {\bf P} \, r_0 (\lambda) \, {\bf P} = 
 r_0 (\lambda) \,, &&
 r_0 (\lambda,\gamma) = r_0 (\lambda,-\gamma) \,, \\
 \label{rg2} 
 (r_0 (\lambda))^t = r_0 (\lambda) \,, &&
 r_0 (\lambda) \, r_0 (-\lambda) = 1\otimes 1 \,,
\end{eqnarray}
where ${\bf P}$ is the permutation in 
$\mathfrak{H} \otimes \mathfrak{H}$ and $t$ denotes
transposition.
The last equality follows from the formula (\ref{rhat}), 
the first relation in (\ref{rprop}) for~$r(\lambda)$,
and the relation $[r(\lambda),\Delta(\varphi_\lambda)]=0$.
Taking logarithmic derivative of (\ref{rg}) at 
$\lambda=0$, we establish the symmetries (\ref{Hh}) 
of the local bulk term $\widehat{H}_{n,n+1}$.
The first relation in (\ref{rg2}) yields in the
same way the second relation in~(\ref{Ht}).

\section{Open chain}
\label{oC}

As we saw in Section~\ref{pR}, $H_{N,1}$ is the
only term in $\cal H$ which does not commute with the
global spin generators ${\cal S}^\pm$. 
Omitting it, we obtain a Hamiltonian
for an {\em open} spin chain,
\begin{equation}\label{Hb}
 {\cal H}^\prime = \sum_{n=1}^{N-1} H_{n,n+1} = 
 \sum_{n=1}^{N-1} \widehat{H}_{n,n+1} + 
 {\rm i} \frac{\sin\gamma}{2} (S^3_1 - S^3_{N}) 
\end{equation}
which is apparently $U_q(sl_2)$-symmetric, i.e.,
$[{\cal H}^\prime, {\cal S}^\pm ]=
 [{\cal H}^\prime, {\cal S}^3 ]=0$. 
It is however not immediately evident whether
the Hamiltonian (\ref{Hb}) remains {\em integrable}.

Let us refer to the sum on the r.h.s.~of~(\ref{Hb}) 
as the bulk Hamiltonian, $\widehat{\cal H}^\prime$.
The remaining part can be called the surface term; 
it is a sum of $(N-1)$ local boundary terms we 
dealt with before. As we discussed above, the bulk 
Hamiltonian $\widehat{\cal H}^\prime$ corresponds to
the $L$-operator (\ref{Lh}) and it is only 
$U(1)$-symmetric. The fact that adding the surface term to 
$\widehat{\cal H}^\prime$ restores the $U_q(sl_2)$-symmetry 
was observed for spin $\frac 12$ in \cite{ABB} and
for spin 1 in \cite{PS}.
Integrability of the corresponding total Hamiltonians
was established in~\cite{bound,Nep1,KS2} in the 
framework of boundary integrable lattice models.  
We will prove below that ${\cal H}^\prime$ is integrable
for the higher spins as well.

Let us briefly recall the construction of an integrable 
Hamiltonian for a chain with boundaries~\cite{bound}.
Let ${\bf \check{R}} (\lambda)$ be a solution of
the Yang-Baxter equation (\ref{RRR}), and $T(\lambda)$
be a monodromy matrix obeying the exchange relation
(\ref{RLL}) with $R$-matrix~${\bf P}{\bf \check{R}}(\lambda)$. 
Introduce a boundary monodromy matrix, 
$Z(\lambda) \equiv T(\lambda) {\cal K}^-(\lambda) 
 (T(-\lambda))^{-1}$.
The boundary matrix,\footnote{%
   Strictly speaking, ${\cal K}^-(\lambda)$ is an element 
   of $\mathfrak{H} \otimes \mathfrak{H}$ but with trivial
   second component. We had a similar situation in
   Section~\ref{tL} where the twist $\varphi_\lambda$ had 
   trivial first component, see eq.~(\ref{tw2}). } 
\hbox{${\cal K}^-(\lambda) \in \mathfrak{H}$}, 
represents non-periodic boundary conditions 
(${\bf \check{R}}$ and ${\cal K}^-$ are close analogues  
of the bulk and boundary scattering matrices~\cite{Sf}).
Now, if ${\cal K}^-(\lambda)$ satisfies the so-called 
reflection equation, namely
\begin{eqnarray}
 \label{refl}
 && {\bf \check{R}} (\lambda-\mu) \, 
   ({\cal K}^-(\lambda) \otimes 1)  \,
 ({\bf \check{R}} (-\lambda-\mu))^{-1} \, 
   (1\otimes  {\cal K}^- (\mu)) \\ 
 \nonumber
 && \qquad\qquad  =  (1\otimes  {\cal K}^- (\mu)) \, 
 {\bf \check{R}}(\lambda+\mu) \, 
  ({\cal K}^-(\lambda) \otimes 1) \,
 ({\bf \check{R}}(-\lambda+\mu))^{-1} \,,
\end{eqnarray}
then $Z(\lambda)$ also satisfies this equation. Using this 
fact one can show that a special trace of $Z(\lambda)$, 
$\tau(\lambda)={\rm tr}{\vphantom |}_0 {\cal K}^+(\lambda)
 Z(\lambda)$, is a generating function for quantum
integrals of motion if the boundary matrix 
${\cal K}^+(\lambda) \in \mathfrak{H}$ satisfies
a ``dual" reflection equation \cite{bound,KuS},
\begin{eqnarray}
 \label{refl2}
 && {\bf \check{R}} (-\lambda+\mu) \, 
   ( ({\cal K}^+)^t (\lambda) \otimes 1)  \,
  {\bf \check{R}} (-\lambda-\mu-2\delta) \, 
   (1\otimes  ({\cal K}^+)^t (\mu)) \\ 
 \nonumber
 && \qquad\qquad  =  (1\otimes  ({\cal K}^+)^t (\mu)) \, 
 {\bf \check{R}}(-\lambda-\mu-2\delta) \, 
  ( ({\cal K}^+)^t (\lambda) \otimes 1) \,
 {\bf \check{R}}(-\lambda+\mu) \,,
\end{eqnarray}
and ${\bf \check{R}}$ has the following properties
\begin{eqnarray}
 \label{uni}
 && {\bf P} \, {\bf \check{R}}(\lambda) \, {\bf P}
 = {\bf \check{R}}(\lambda) \,, \quad
 {\bf \check{R}}(\lambda) \, {\bf \check{R}}(-\lambda)
 = \rho_1(\lambda) \, (1 \otimes 1) \,, \quad
 \bigl({\bf \check{R}}(\lambda)\bigr)^t =
 {\bf \check{R}}(\lambda) \,, \\
 \label{cro}
 && ( {\bf \check{R}}(\lambda) )^{t_1} \, 
 ( {\bf \check{R}}(-\lambda - 2\delta ) )^{t_1} =
 \rho_2(\lambda) \, (1 \otimes 1) \,,
\end{eqnarray}
where $\delta$ is a constant, and
$\rho_1(\lambda)$ and $\rho_2(\lambda)$ are scalar function.

With all these conditions, an integrable Hamiltonian 
is given \cite{bound} by the following analogue of 
the formula~(\ref{Hq2}) (we keep the same normalization
as in (\ref{Hq2}))
\begin{eqnarray}
 \label{Hbo}
 {\cal H}^{\prime\prime} 
 &=&   \sum_{n=1}^{N-1} h_{n,n+1} 
 + {\rm i} \frac{\sin\gamma}{4\gamma} \, 
   \frac{d}{d\lambda} {\cal K}^-_1(\lambda)
  \Bigl\vert_{\lambda=0}  \Bigr. 
 + \frac{  {\rm tr}{\vphantom|}_0 
      \bigl( {\cal K}^+_0 (0) \, h_{0,N} \bigr) }{
    {\rm tr}\, {\cal K}^+ (0)  }  \,,\\
 \nonumber
 && h_{n,n+1} = {\rm i} \frac{\sin\gamma}{2\gamma} \, 
 {\bf P}_{n,n+1} \, 
 \frac{d}{d\lambda} {\bf \check{R}}_{n,n+1}(\lambda)
 \Bigl\vert_{\lambda=0}  \Bigr.  
\end{eqnarray}
Deriving (\ref{Hbo}) one assumes that 
${\bf \check{R}}_{n,n+1}(0)= {\bf P}$ and 
${\cal K}^-(0) = 1 $, which is consistent with~(\ref{refl}).

Let us try to identify the Hamiltonian (\ref{Hb}) as a 
particular case of (\ref{Hbo}). First, we can put
$h_{n,n+1} = \widehat{H}_{n,n+1}$ if we choose 
${\bf \check{R}}(\lambda)={\bf \check{R}}_\gamma(\lambda)$,
where ${\bf \check{R}}_\gamma(\lambda)=
 r_0(\lambda) {\bf P}$ is given by 
(\ref{Rhat}) with $\beta=\gamma$. For this $R$-matrix
the properties (\ref{uni}) follow from~(\ref{rg});
in particular, the second relation (unitarity) holds 
with $\rho_1(\lambda)=1$. The crossing unitarity (\ref{cro})
holds for ${\bf \check{R}}_\gamma(\lambda)$ with 
$\delta= {\rm i}$.

The derivation of the reflection equation (\ref{refl})
for ${\cal K}^-$ does not use the conditions 
(\ref{uni})-(\ref{cro}). So, let us look first for an 
$R$-matrix for which the reflection equation has a 
trivial solution, ${\cal K}^-(\lambda) = 1$. 
In this case (\ref{refl}) turns into
\begin{equation} \label{ref0}
 {\bf \check{R}} (\lambda-\mu) \, 
 ({\bf \check{R}} (-\lambda-\mu))^{-1} \, 
  =  {\bf \check{R}}(\lambda+\mu) \, 
 ({\bf \check{R}}(-\lambda+\mu))^{-1} \,.
\end{equation}
A solution to this equation is given by 
${\bf \check{R}} (\lambda) = r(\lambda) {\bf P}$, where
$r(\lambda)$ is the universal $r$-matrix we discussed in
Section~\ref{uR}. Indeed, (\ref{ref0})
follows {}from the second relation in~(\ref{rprop}). 
Using this observation, we can find a solution to
the reflection equation for any $R$-matrix 
${\bf \check{R}}_\beta(\lambda)$ given by~(\ref{Rhat}). 
Indeed, substituting (\ref{Rhat}) in (\ref{ref0}), 
we derive
\begin{eqnarray}
 \label{refb}
 && {\bf \check{R}}_\beta (\lambda-\mu) \, 
   (1 \otimes \varphi_\lambda^2)  \,
 ({\bf \check{R}}_\beta (-\lambda-\mu))^{-1} \, 
   (1 \otimes \varphi_{-\mu}^2) \\ 
 \nonumber
 && \qquad\qquad  = (1 \otimes \varphi_{-\mu}^2) \, 
 {\bf \check{R}}_\beta (\lambda+\mu) \, 
  (1 \otimes \varphi_\lambda^2) \,
 ({\bf \check{R}}_\beta (-\lambda+\mu))^{-1} \,.
\end{eqnarray}
Multiplying this relation by 
$\Delta(\varphi_\lambda^{-2}) =
 (\varphi_\lambda \otimes \varphi_\lambda)^{-2}$ (which
commutes with ${\bf \check{R}}_\beta (\lambda)$), we
bring it to the form of~(\ref{refl}). Thus, a solution
of the reflection equation for ${\bf \check{R}}_\beta$ is
\[  
 {\cal K}_\beta^-(\lambda) = (\varphi_\lambda)^{-2} = 
 e^{-2\beta\lambda S^3} \,.
\]

Notice that for $\beta=\gamma$ the $R$-matrix 
${\bf \check{R}}$ is symmetric with respect to $\gamma$ 
(the second relation in (\ref{rg})). Therefore in this 
case we have another solution (which can be derived 
directly {}from (\ref{refb}) by applying the permutation and 
then multiplying by $(\varphi_\mu \otimes \varphi_\mu)^2$),
\begin{equation}\label{Ku2}
 {\cal K}^-(\lambda) = e^{2\gamma\lambda S^3} \,.
\end{equation}
This is the boundary matrix we need since
its derivative in (\ref{Hbo})
gives exactly the $S_1^3$ term in~(\ref{Hb}).

In general, solutions of the reflection equation
and of its ``dual'' are independent of each other.
However, as was noted already in \cite{bound}, there
exist several isomorphisms that allow us to construct
${\cal K}^+$ if we know ${\cal K}^-$. In particular, 
it is easy to see that a possible solution for 
(\ref{refl2}) is
$
 {\cal K}^+ (\lambda) = ({\cal K}^- (-\lambda -\delta) )^t 
$.
For ${\cal K}^-$ given by (\ref{Ku2}) this yields
\begin{equation}\label{Kpg}
  {\cal K}^+ (\lambda) = 
  e^{-2\gamma(\lambda + {\rm i}) S^3} \,. 
\end{equation}
It turns out that substitution of 
$ {\cal K}^+ (0) = q^{-2 S^3}$ into (\ref{Hbo})
gives exactly the $S_N^3$ term in~(\ref{Hb}). To
prove this assertion, we first observe (see 
Appendix~C) that for the $q$-trace of the local 
Hamiltonian $H_{n,n+1}$ we have
\begin{equation}\label{qtr}
  {\rm tr}{\vphantom|}_n 
      \bigl(  q^{2 S_n^3} \, H_{n,n+1} \bigr)
  = \tilde{\rho}_S \, 1_{n+1} \,, 
\end{equation}
where $\tilde{\rho}_S$ is a scalar constant.

Taking into account the relation (\ref{dec})
between $H_{n,n+1}$ and $\widehat{H}_{n,n+1}$,
we infer from (\ref{qtr}) that 
$ {\rm tr}{\vphantom|}_0 
 \bigl(  q^{2 S_0^3} \, \widehat{H}_{0,N} \bigr)
  = ({\rm tr}\, q^{2 S^3}) \, {\rm i} 
 \frac{\sin\gamma}{2} \, S_N^3 +  
  {\rm const}$. Replacing now $q$ with
$q^{-1}$ and using that $\widehat{H}_{n,n+1}$
is an even function of $\gamma$ (\ref{Hh}), we
find that 
\begin{equation}\label{Kptr}
 \frac{ {\rm tr}{\vphantom|}_0 \bigl( {\cal K}^+ (0) \,
 \widehat{H}_{0,N} \bigr) }{ {\rm tr}\, {\cal K}^+ (0) } 
 = \frac{ {\rm tr}{\vphantom|}_0 \bigl( q^{-2 S_0^3} \,
 \widehat{H}_{0,N} \bigr) }{ {\rm tr}\, q^{-2 S^3} }
  = - {\rm i} \frac{\sin\gamma}{2} S_N^3
\end{equation}
holds (up to an additive constant). Thus, the boundary
matrix (\ref{Kpg}) gives the $S_N^3$ term 
in~(\ref{Hb}). This completes the proof that the 
$U_q(sl_2)$-symmetric open chain Hamiltonian (\ref{Hb}) 
is integrable for all (half-) integer spins.

\section*{Conclusion}
In summary, we have constructed explicitly (in terms
of the spin generators) higher spin closed chain 
Hamiltonians for two families of $XXZ$-type $L$-operators 
including the two \hbox{$L$-operators} most often used 
in the literature. We have investigated properties of 
these Hamiltonians, described their interrelations, 
and discussed the connection with 
$U_q(sl_2)$-symmetric open chain Hamiltonians.

We have emphasized a key r{\^o}le of the underlying
quantum algebraic structure, especially the universal 
$r$-matrix and the co-multiplication, for 
constructing integrable Hamiltonians and investigating 
their symmetries. The technique presented in this paper 
can be applied also for constructing the higher 
quantum integrals of motion.
 
Our construction does not in general extend to the
case when $q$ is a root of unity. 
However it appears that we could allow $q^p=1$ if $p$ 
is sufficiently large in comparison with the spin~$S$. 
For instance, examining the denominator of (\ref{Pj})
we see that the projectors ${\cal P}_j$ do not 
become singular at roots of unity if~$p>8S$. 
If the ratio $p/S$ is not large, then an appropriate 
modification of the presented construction is needed; 
it will be certainly of interest and use.

\vspace*{2mm}
\par\noindent {\bf Acknowledgments:}
I am grateful to H.M.~Babujian for useful discussions
and to L.D.~Faddeev, P.P.~Kulish, and J.~Teschner for 
valuable comments.
This work was supported by Alexander von Humboldt 
Foundation and in part by grants INTAS 99-01459
and RFBR 99-01-00101.

\section*{Appendix A}
Here we give some technical details on the universal
$r$-matrices used in the text. First we
recall the derivation of the universal $r$-matrix 
(\ref{unr}) along the lines of \cite{Ji1,Ji2} (see also
 \cite{Fad,De}). The explicit form of the off-diagonal 
entries of equation (\ref{rll0}) is
\begin{equation}\label{rS}
 r(\lambda) \, (e^{\mp \gamma\lambda} 
 S^\pm \otimes q^{S^3} +  e^{\pm \gamma\lambda} 
 q^{-S^3} \otimes S^\pm)  =
 (e^{\pm \gamma\lambda} S^\pm \otimes q^{S^3} + 
 e^{\mp \gamma\lambda} q^{-S^3} \otimes S^\pm) \, 
 r(\lambda) \,.  
\end{equation}
A solution to this equation is unique up to a scalar 
factor \cite{Ji2}. Notice that the co-multiplication 
(\ref{del}) satisfies the following relations
\[
 [ S^\varepsilon \otimes q^{S^3}, \Delta(S^\varepsilon) ] =
 [ q^{-S^3} \otimes S^\varepsilon , \Delta(S^\varepsilon) ] = 0 \,,
 \qquad \varepsilon = \pm \,.
\]
Therefore, for the highest/lowest weight vectors we 
deduce that
\begin{equation}\label{jj}
 S^\pm \otimes q^{S^3} \, |j, \pm j \rangle = 
 g_\pm(j) |j+1, \pm (j+1) \rangle \,, \qquad
 q^{-S^3} \otimes S^\pm \, |j, \pm j \rangle = 
 h_\pm(j) |j+1, \pm (j+1) \rangle \,, 
\end{equation}
where $g_\pm(j)$ and $h_\pm(j)$ are scalar functions. 
Furthermore, we have
\begin{eqnarray}
 \label{jjj}
 && ( S^\pm \otimes q^{S^3} + q^{\pm(2+2j)}
 q^{-S^3} \otimes S^\pm) \, |j,\pm j \rangle =
 \Bigl( (1\otimes q^{2S^3}) \, S^\pm \otimes q^{-S^3} \\
 && + q^{\pm(2+2j)} (q^{S^3} \otimes S^\pm q^{2S^3})
 q^{-2S^3} \otimes q^{-2S^3} \Bigr) \, |j, \pm j \rangle 
 =  (1\otimes q^{2S^3}) \, \Delta (S^\pm) \, |j, \pm j \rangle
 = 0 \,.  \nonumber
\end{eqnarray}
Applying now (\ref{rS}) to $|j,j \rangle$ (for the upper signs)
or to $|j, -j \rangle$ (for the lower signs) and using 
(\ref{rP}), (\ref{jj}) and (\ref{jjj}), we deduce that
\begin{eqnarray}
 \label{rjja}
 & r_{j+1}(\lambda) \, (e^{\pm \gamma\lambda} -
 q^{\pm(2+2j)} e^{\mp \gamma\lambda}) =
 (e^{\mp \gamma\lambda} - q^{\pm(2+2j)} e^{\pm \gamma\lambda})
 \ r_{j}(\lambda) \,. &
\end{eqnarray}
Both relations in (\ref{rjja}) yield the same functional 
equation on $r_{j}(\lambda)$ (which arises also for 
universal $r$-matrices in the lattice sine-Gordon model 
\cite{Tar} and in the lattice Virasoro algebra~\cite{Vir})
\begin{equation}\label{rj}
 r_{j+1}(\lambda) = \frac{\sin[\gamma(j+1-i\lambda)]}%
 {\sin[\gamma(j+1+i\lambda)]}  \  r_{j}(\lambda)  \,.
\end{equation}
Upon imposing the normalization condition $r(0)=1$,
we obtain the expression (\ref{unr}). A proof that 
this $r(\lambda)$ does satisfy all the relations 
in (\ref{rll0}) is given in \cite{Ji1,Ji2}.
Of course, one is still free to multiply $r(\lambda)$
by a scalar function, $\rho(\lambda)$, such that
$\rho(0)=1$. For instance, the spin-$\frac 12$ 
representation of $\check{R}(\lambda)$ given by 
(\ref{R}) corresponds to
$\rho (\lambda) = \sin[\gamma(1+i\lambda)]/\sin\gamma$.

Consider now the universal $r$-matrix $r_0(\lambda)$
introduced in Section~\ref{tL}. It is a solution to 
equation (\ref{rLL}) for the $L$-operator (\ref{Lh}). 
Unlike the previous case, this $L$-operator does not 
have a Borel decomposition of the type (\ref{Bor}).
Therefore the off-diagonal entries of (\ref{rLL}) 
give us four $\lambda$-dependent relations:
\begin{eqnarray}
 \label{rhSa}
 && r_0(\lambda) \, ( S^\pm \otimes q^{S^3} +  
 e^{\pm \gamma\lambda} q^{-S^3} \otimes S^\pm)  =
 (e^{\pm \gamma\lambda} S^\pm \otimes q^{S^3} + 
 q^{-S^3} \otimes S^\pm) \, r_0(\lambda) \,, \\ 
 \label{rhSb}
 && r_0(\lambda) \, ( S^\pm \otimes q^{-S^3}  
 e^{\pm \gamma\lambda} + q^{S^3} \otimes S^\pm)  =
 (e^{\pm \gamma\lambda} q^{S^3} \otimes S^\pm + 
 S^\pm \otimes q^{-S^3} ) \, r_0(\lambda) \,. 
\end{eqnarray}
It can be proven along the lines of \cite{Ji2} that
solution to this equation is unique up to a scalar 
factor. As we have already shown in Section~\ref{tL},
this solution is given by  $r_0(\lambda) = 
 (1 \otimes e^{\gamma \lambda S^3}) r(\lambda) 
 (e^{-\gamma \lambda S^3} \otimes 1)$,
where $r(\lambda)$ solves (\ref{rS}). 

Now we observe that 
$r_1(\lambda) \equiv {\bf P} r_0(\lambda) {\bf P}$ and 
$r_2(\lambda) \equiv (r_0(\lambda))^t $
solve the same set of equations (\ref{rhSa})-(\ref{rhSb}).
By the above mentioned uniqueness, this implies
that $r_i(\lambda) = c_i(\lambda) r_0(\lambda)$,
where $c_i(\lambda)$, $i=1,2$ are scalar functions.
Since ${\bf P}^2 = 1\otimes 1$, and $((r_0)^t)^t=r_0$, 
we conclude that $c_i(\lambda) = \pm 1$. 
Imposing the condition $r_0(0)= 1 \otimes 1$, we 
have to put $c_i(\lambda) = 1$.
Thus, we have proven the relations on the l.h.s~of 
(\ref{rg}) and (\ref{rg2}). Employing the first of 
them and the property (\ref{rPg}) of $r(\lambda)$, 
we prove the second relation in (\ref{rg}) as follows:
\begin{eqnarray*}
 r_0 (\lambda,-\gamma) &=& 
 (1 \otimes e^{-\gamma \lambda S^3}) r(\lambda,-\gamma) 
 (e^{\gamma \lambda S^3} \otimes 1) =
 (1 \otimes e^{-\gamma \lambda S^3}) {\bf P} r(\lambda,\gamma) 
 {\bf P} (e^{\gamma \lambda S^3} \otimes 1) \\
 &=& {\bf P} (e^{-\gamma \lambda S^3} \otimes 1) 
 r(\lambda,\gamma) (1 \otimes e^{\gamma \lambda S^3}) {\bf P} =
 {\bf P} (1 \otimes e^{\gamma \lambda S^3})
 r(\lambda,\gamma) (e^{-\gamma \lambda S^3} \otimes 1) {\bf P} \\
 &=&  {\bf P} r_0 (\lambda,\gamma) {\bf P} =
 r_0 (\lambda,\gamma) \,.
\end{eqnarray*}
In the second line we used that 
$[r(\lambda),\Delta(e^{\gamma\lambda S^3})]=0$.

\section*{Appendix B}
Here we provide some details on computation of the 
Hamiltonians in the cases of spin $1$ and spin $\frac 32$.
For $S=1$, a matrix representation of the spin generators is
(as was discussed in Section~\ref{pR}, one has a freedom of
rescaling $S^\pm \rightarrow \eta^{\mp \frac 12} S^\pm$
with any real $\eta \neq 0$)
\begin{equation}\label{S1z}
 S^+ =  \left( 
   \begin{array}{ccc} 0 & a & 0 \\ 0 & 0 & a \\ 0 & 0 & 0
   \end{array} \right)   \,, \quad
 S^- =  (S^+)^t  , \quad
 S^3 =  \left(
   \begin{array}{ccc} 1 & 0 & 0 \\ 0 & 0 & 0 \\ 0 & 0 & -1
   \end{array} \right)  \,, 
\end{equation}
where $t$ denotes transposition and $a=\sqrt{2\cos\gamma}$.
Since $(S^3)^3=S^3$, any function of $S^3$ is a polynomial
in $S^3$ of a degree not exceeding two. In particular, we have
\begin{equation}\label{s1a}
 \sin(t S^3) = S^3 \sin t \,,\qquad
 \cos(t S^3) = 1 - 2 (S^3)^2 \sin^2\frac{t}{2} \,.
\end{equation}
Denote $\Upsilon = \frac 12 
  (S^+_{n} S^-_{n+1} + S^-_{n} S^+_{n+1})$.
With the help of formulae (\ref{s1a}) we rewrite 
(\ref{Hnn}) as follows
\begin{eqnarray}
 \label{x1a}
 {\bf X}_{1} &=& q^{S^3_n} \, \Upsilon \, q^{-S^3_{n+1}}
 + (\cos\gamma) \, \Bigl( \Lambda + 
     \frac 12 + 2 (\cos\gamma)^2 \Bigr) \,, \\
 \nonumber
 \Lambda &=&
 \frac 32 - 2 (\cos\gamma)^2 + 
 (\cos\gamma)^2 (S^3_{n} S^3_{n+1})  
 +  (\sin\gamma)^2 \Bigl( (S^3_{n} S^3_{n+1})^2  
 - 2 (S^3_{n})^2 - 2 (S^3_{n+1})^2 \Bigr) \\
 \nonumber
 && +\,  {\rm i} (\sin 2\gamma) \Bigl( S^3_{n} - S^3_{n+1}
 + \frac 12 (S^3_{n})^2 S^3_{n+1} 
 - \frac 12 S^3_{n} (S^3_{n+1})^2 \Bigr)   \,.
\end{eqnarray}
Substituting (\ref{x1a}) into (\ref{H1a}), 
applying several times formulae (\ref{s1a}) and (\ref{Cn1}), 
we obtain the Hamiltonian in the following form 
\begin{eqnarray}
 \label{H1S}
 H_{n,n+1} &=& \frac{1}{4 \cos\gamma} 
 \Bigl( 
  - \frac{1}{\cos^2\gamma} (\Upsilon)^2  
  - \frac{1}{\cos\gamma} 
 q^{S^3_n} \, (\Upsilon \, \Lambda +
 \Lambda \, \Upsilon )\, q^{-S^3_{n+1}}  \Bigr.  \\
 \nonumber
 &&  +\, 4 + 2\cos 2\gamma +
  (\cos 2\gamma) \bigl(S^3_{n} S^3_{n+1}
  - (S^3_{n} S^3_{n+1})^2 \bigr)  \\
 \nonumber
 && -\, 2 (\sin\gamma)^2 \bigl( (S^3_{n})^2 + (S^3_{n+1})^2 \bigr)
   \Bigl. \Bigr) +
 {\rm i} \frac{\sin\gamma}{2} \, ( S^3_{n} - S^3_{n+1})  \,.
\end{eqnarray}
Deriving the first term here we used the following 
analogue of relation (\ref{ex1})
\begin{equation}\label{id0}
 e^{t S^3} \, (S^\pm)^2  = (S^\pm)^2 \, e^{-t S^3} = 
 e^{\pm t } \, (S^\pm)^2 \,.    
\end{equation}
Next, we observe that the following identity holds
(as can be checked directly in terms of matrices)
\begin{equation}\label{id2}
 - F_\alpha \, \bigl( q^{S^3_n} \, 
 (\Upsilon \, \Lambda + \Lambda \, \Upsilon ) \, 
 q^{-S^3_{n+1}} \bigr) \, (F_\alpha)^{-1}
 = \Theta^-_\alpha \, \Upsilon 
  + \Upsilon \, \Theta^+_\alpha \,,
\end{equation}
with $F_\alpha$ as in (\ref{twi1}) and
\begin{equation}\label{Q2}
 \Theta^\pm_\alpha = \frac 12 (1\otimes 1) +
 (1-2 q^{\pm \alpha} \cos\gamma) \, S^3_{n} S^3_{n+1} \,.
\end{equation}
Substituting (\ref{id2})-(\ref{Q2}) with $\alpha=0$
into (\ref{H1S}), we obtain the Hamiltonian (\ref{H1Sb}).

The Hamiltonian $\tilde{H}_{n,n+1}^{(\alpha)}$
discussed in Section~\ref{tC} is
obtained from (\ref{H1S}) by the twist~(\ref{Ha}).
Notice that the term $(\Upsilon)^2$ in (\ref{H1S})
is not affected due to (\ref{id0}). The diagonal
part of Hamiltonian apparently commutes with the
twist. So the only part of (\ref{H1S}) which 
changes is the second term. It transforms  
according to (\ref{id2}), which yields the 
Hamiltonian (\ref{H1Sc}).

Consider now the case $S=\frac 32$. The spin generators
are given by
\begin{equation}\label{S32z}
 S^+ =  \left( 
   \begin{array}{cccc} 0 & a_1 & 0 & 0 \\ 0 & 0 & a_2 & 0 \\ 
   0 & 0 & 0 & a_1 \\ 0 & 0 & 0 & 0
   \end{array} \right)   , \quad
 S^- = (S^+)^t \,, \quad
 S^3 =  \left(
   \begin{array}{cccc} \frac 32 & 0 & 0 & 0 \\ 
 0 & \frac 12 & 0 & 0 \\ 0 & 0 & - \frac 12 & 0 \\
 0 & 0 & 0 & -\frac 32  \end{array} \right)  , 
\end{equation}
with $a_1=(2\cos2\gamma +1)^{\frac 12}$ and $a_2=2\cos\gamma$.
Any function of $S^3$ is a polynomial in $S^3$
of a degree not exceeding three. In particular,
\begin{eqnarray}
 \label{ex3a}
 \sin(t S^3) &=& (2 \sin \frac t2 + \frac 13 \sin^3 \frac t2 )\,
 S^3 - \frac 43 (\sin^3 \frac t2) \, (S^3)^3  \,, \\
  \label{ex3b}
 \cos(t S^3) &=&  \cos \frac t2 + 
 \frac 14 \sin \frac t2 \sin t - 
 (\sin \frac t2 \sin t) \, (S^3)^2 \,.
\end{eqnarray}

The Hamiltonian (\ref{H32a}) for $S=\frac 32$ takes
the following form in terms of ${\bf X}_{\frac 32}$
\begin{eqnarray}
 \label{H32b} 
 H_{n,n+1} &=&  
 \frac{1}{4 (\cos\gamma)^3 (1+2\cos 2\gamma)^3} 
 \Bigl( 4 (\cos\gamma) \, ({\bf X}_{\frac 32})^2 \\
 \nonumber 
 && -\, ({\bf X}_{\frac 32})^2 \,
 ( 13 + 20 \cos 2\gamma + 8 \cos 4\gamma 
 + 2 \cos 6\gamma ) \\
 \nonumber
 && +\, {\bf X}_{\frac 32} \, 
 ( 68 \cos\gamma + 48 \cos 3\gamma + 23 \cos 5\gamma 
 + 7 \cos 7\gamma + \cos 9\gamma ) \Bigr) \,.
\end{eqnarray}
This Hamiltonian can be rewritten in the same way  
as we did for $S=1$ using, in particular, relations 
(\ref{Cn32}), (\ref{ex3a}) and (\ref{ex3b}), and 
appropriate analogues of (\ref{id0}) and (\ref{id2}). 
The final form is given by (\ref{H32Sb}), with
\begin{eqnarray}
 \nonumber 
 {\bf Q} &=&  \frac{1}{192} \, (1\otimes 1) \,
 \bigl(815 - 5832 \cos\gamma + 2352 \cos 2\gamma 
  - 3888 \cos 3\gamma \\
 \nonumber 
 && \qquad  + 2542\cos 4\gamma 
 - 1620\cos 5\gamma + 1600\cos6\gamma - 
        324\cos 7\gamma + 467\cos 8\gamma \bigr) \\
 \nonumber 
 &&  + \frac 23 (\sin \frac{\gamma}{2})^2 \, S^3_n S^3_{n+1} \,
 \bigl( 106 + 37 \cos\gamma + 186 \cos 2\gamma 
 + 29 \cos 3\gamma \\
 \nonumber 
 && \qquad  + 88 \cos 4\gamma + 13 \cos 5\gamma + 
    28 \cos 6\gamma + 5 \cos 7\gamma \bigr) \\
 \nonumber
 &&  + (\sin \frac{\gamma}{2})^2 \, (S^3_n S^3_{n+1})^2 \,
 \bigl( 28 - 53 \cos\gamma + 54 \cos 2\gamma 
 - 31 \cos 3\gamma \\
 \nonumber 
 && \qquad  + 28 \cos 4\gamma - 11 \cos 5\gamma 
 + 10 \cos 6\gamma - \cos 7\gamma \bigr) \\
 \nonumber
 && + \frac{1}{12} (\sin \frac{\gamma}{2})^2 \,
 \Bigl( (S^3_n)^2 + (S^3_{n+1})^2 \Bigr) \,
 \bigl( 412 + 1687 \cos\gamma + 606 \cos 2\gamma  \\
 \nonumber 
 && \qquad  + 1085 \cos 3\gamma + 268 \cos 4\gamma 
 + 433 \cos 5\gamma 
 + 58 \cos 6\gamma + 83 \cos 7\gamma \bigr) \,, \\[0.5mm]
 \nonumber  
 {\bf F}_{\frac 32} &=& \frac{\cos^2\gamma}{4} \, (1\otimes 1) \,
 \bigl( 10328 + 17080 \cos 2\gamma + 9071 \cos 4\gamma 
 + 3220 \cos 6\gamma + 621 \cos 8\gamma \bigr) \\
 \nonumber
 && - \frac 19 (\sin\frac{\gamma}{2})^2 \, S^3_n S^3_{n+1} \,
 \bigl( 8364 + 5020 \cos\gamma + 12752 \cos 2\gamma 
 + 2499 \cos 3\gamma + 7150 \cos 4\gamma \\
 \nonumber
 && \qquad + 1023 \cos 5\gamma + 3320 \cos 6\gamma 
 + 447 \cos 7\gamma + 814 \cos 8\gamma + 83 \cos 9\gamma \bigr) \\
 \nonumber
 && - \frac 43 (\sin\frac{\gamma}{2})^2 \, (S^3_n S^3_{n+1})^2 \,
 \bigl(536 - 472 \cos\gamma + 864 \cos 2\gamma 
 - 347 \cos 3\gamma + 650 \cos 4\gamma \\
 \nonumber
 && \qquad - 163 \cos 5\gamma + 272 \cos 6\gamma 
 - 65 \cos 7\gamma + 78 \cos 8\gamma - 9 \cos 9\gamma \bigr) \\
 \nonumber
 && - \frac{64}{9} (\sin\frac{\gamma}{2})^4 \, 
 (S^3_n S^3_{n+1})^3 \,
 \bigl( 50 - 200 \cos\gamma + 48 \cos 2\gamma - 168 \cos 3\gamma 
 + 45 \cos 4\gamma \\
 \nonumber
 && \qquad - 76 \cos 5\gamma + 28 \cos 6\gamma 
 - 20 \cos 7\gamma + 13 \cos 8\gamma \bigr) \\
 \nonumber
 && - (\sin 2\gamma)^2 \, \Bigl( (S^3_n)^2 + (S^3_{n+1})^2 \Bigr)
 \, \bigl( 169 + 274 \cos 2\gamma 
 + 106 \cos 4\gamma + 27 \cos 6\gamma \bigr) \\
 \nonumber
 && + \frac 49 (\sin\frac{\gamma}{2})^2 \,
 \Bigl( (S^3_n)^3 S^3_{n+1} + S^3_{n} (S^3_{n+1})^2 \Bigr) \,
 \bigl(444 - 2516 \cos\gamma 
 + 560 \cos 2\gamma - 2037 \cos 3\gamma \\
 \nonumber
 && \qquad  + 334 \cos 4\gamma 
 - 1137 \cos 5\gamma + 200 \cos 6\gamma 
 - 417 \cos 7\gamma + 46 \cos 8\gamma - 85 \cos 9\gamma \bigr) \,.
\end{eqnarray}
The equivalence of (\ref{H32b}) and (\ref{H32Sb})
as 16$\times$16 matrices has been verified with the 
help of the program Mathematica.

\section*{Appendix C}
Here we explain the origin of the equation (\ref{qtr}) 
that was important for our discussion on the open 
chain Hamiltonian in Section~\ref{oC}.

Recall that, by definition (see e.g., \cite{twi1,Dr2}), 
a Hopf algebra ${\cal A}$ possesses the antipode map
$s : {\cal A} \rightarrow {\cal A}$
which is an anti-homomorphism consistent with the
co-multiplication and the co-unit in the sense that
$m((s\otimes id) \Delta(\xi)) = 
 m((id \otimes s) \Delta(\xi)) = 
 \epsilon(\xi) \cdot 1$ (here 
 $m : {\cal A}^{\otimes 2} \rightarrow {\cal A}$
is the multiplication).
Assume that there exists an element $\chi \in {\cal A}$ 
which realizes the square of the antipode (which 
is a homomorphism) as an inner automorphism, i.e.,
$ s(s(\xi)) = \chi \, \xi \,  \chi^{-1} $
for any $\xi \in {\cal A}$. Then, as was
proven in \cite{Dr2}, 
\begin{equation}\label{trc}
 {\rm tr}{\vphantom |}_1 \bigl( 
 ( \chi^{-1} \otimes 1) \, b \bigr)
\end{equation}
belongs to the center of ${\cal A}$ if
an element $b \in {\cal A}^{\otimes 2}$
satisfies $[b,\Delta(\xi)] = 0$ 
for any $\xi \in {\cal A}$.

For ${\cal A} = U_q(sl_2)$ the antipode 
consistent with the co-multiplication 
(\ref{del}) is given by 
\[
 s(S^\pm) = - q^{\mp 1} \, S^\pm \,,
 \qquad s(S^3) = - S^3 \,.
\]
It is easy to see that in this case 
$\chi = q^{-2 S^3}$. Since the universal 
$r$-matrix $r(\lambda)$ has the property 
(\ref{rd}), we can apply (\ref{trc}) and
infer that its $q$-trace,
${\rm tr}{\vphantom |}_1 \bigl( 
 ( q^{2 S^3} \otimes 1) \, r(\lambda) \bigr)$,
belongs to the center of~$U_q(sl_2)$.
Consequently, the same holds for the local 
Hamiltonian $H_{n,n+1}$ constructed as in~(\ref{Hq2}). 
Being evaluated in an irreducible representation,
the $q$-trace of $H_{n,n+1}$ becomes just 
a constant, as was stated in eq.~(\ref{qtr}).

\end{document}